\documentclass[11pt]{article}   	
\usepackage{amsmath,amssymb,amsfonts,mathrsfs,bbm,bm,slashed}
\usepackage{mathrsfs,mathtools}
\renewcommand\mathbb[1]{\mathbbm{#1}}

\usepackage{tocloft} % well-formatted table of contents
\usepackage[nosort]{cite} % better citation format
\usepackage[usenames,dvipsnames]{xcolor}
\usepackage{graphicx,color}
\usepackage[colorlinks=true, linkcolor=blue, citecolor=blue, linktoc=all]{hyperref}
\usepackage{tikz-cd}
\usepackage{pict2e}
\usepackage{physics}
\usepackage{comment} 
\usepackage[makeroom]{cancel}

\newcommand{\thistitle}{\huge BRST Cohomology is Lie Algebroid Cohomology}
\newcommand{\addressuiuc}[1]{
	\centerline{
		\begin{minipage}[c]{0.7\textwidth}
			\begin{center}
			${}^{#1}$ Illinois Center for Advanced Studies of the Universe \& Department of Physics,\\ 
			University of Illinois, 1110 West Green St., Urbana IL 61801, U.S.A.
			\end{center}
		\end{minipage}
		}
	}

\newcommand{\beq}{\begin{eqnarray}}
\newcommand{\eeq}{\end{eqnarray}}
\newcommand{\beqn}{\begin{eqnarray}}
\newcommand{\eeqn}{\end{eqnarray}}
\newcommand{\bea}{\begin{eqnarray}}
\newcommand{\eea}{\end{eqnarray}}
\newcommand{\be}{\begin{equation}}
\newcommand{\ee}{\end{equation}}
\newcommand{\un}[1]{\underline{#1}}
\def\pa{\partial}
\usepackage[normalem]{ulem}

\newcommand{\RR}{\mathbb{R}}

\newcommand{\hlt}[1]{{\color{Emerald}{\em #1}}}

\newcommand{\nn}{\nonumber}

\newcommand{\scr}{\mathscr}
\renewcommand{\tr}{\text{tr}}
\newcommand{\td}{{\rm d}}
\newcommand{\ts}{{\rm s}}

\newcommand{\mX}{\mathfrak{X}}
\newcommand{\mY}{\mathfrak{Y}}

\newcommand{\mg}{\mathfrak{g}}

\newcommand{\End}{\text{End}}
\newcommand{\Der}{\text{Der}}
\newcommand{\hatd}{\hat{\td}}

\newcommand{\Aconn}[1]{\phi_{#1}}

\newcommand{\econn}{\hat{A}}
\newcommand{\ecurv}{\hat{F}}

\DeclareMathAlphabet{\mymathbb}{U}{BOONDOX-ds}{m}{n}

\makeatletter
\DeclareRobustCommand{\loplus}{\mathbin{\mathpalette\dog@lsemi{+}}}
\DeclareRobustCommand{\lotimes}{\mathbin{\mathpalette\dog@lsemi{\times}}}
\DeclareRobustCommand{\roplus}{\mathbin{\mathpalette\dog@rsemi{+}}}
\DeclareRobustCommand{\rotimes}{\mathbin{\mathpalette\dog@rsemi{\times}}}

\newcommand{\dog@rsemi}[2]{\dog@semi{#1}{#2}{-90,90}}
\newcommand{\dog@lsemi}[2]{\dog@semi{#1}{#2}{270,90}}
\newcommand{\dog@semi}[3]{%
  \begingroup
  \sbox\z@{$\m@th#1#2$}%
  \setlength{\unitlength}{\dimexpr\ht\z@+\dp\z@\relax}%
  \makebox[\wd\z@]{\raisebox{-\dp\z@}{%
    \begin{picture}(1,1)
    \linethickness{\variable@rule{#1}}
    \roundcap
    \put(0.5,0.5){\makebox(0,0){\raisebox{\dp\z@}{$\m@th#1#2$}}}
    \put(0.5,0.5){\arc[#3]{0.5}}
    \end{picture}%
  }}%
  \endgroup
}
\newcommand{\variable@rule}[1]{%
  \fontdimen8  
  \ifx#1\displaystyle\textfont3\else
    \ifx#1\textstyle\textfont3\else
      \ifx#1\scriptstyle\scriptfont3\else
        \scriptscriptfont3\relax
  \fi\fi\fi
}
\DeclareRobustCommand{\loplus}{\mathbin{\mathpalette\dog@lsemi{+}}}

\begin{document}

\title{\thistitle}
\author{Weizhen Jia\footnote{weizhen2@illinois.edu},\,
	    Marc S.~Klinger\footnote{marck3@illinois.edu}\, and 
	    Robert G.~Leigh\footnote{rgleigh@illinois.edu}
	\\
	\\ 
	{\small \emph{\addressuiuc{}}} 
	}
\date{}
\maketitle

\begin{abstract}

In this paper we demonstrate that the exterior algebra of an Atiyah Lie algebroid generalizes the familiar notions of the physicist's BRST complex. To reach this conclusion, we develop a general picture of Lie algebroid isomorphisms as commutative diagrams between algebroids preserving the geometric structure encoded in their brackets. We illustrate that a necessary and sufficient condition for such a diagram to define a morphism of Lie algebroid brackets is that the two algebroids possess gauge-equivalent connections. This observation indicates that the aforementioned set of Lie algebroid isomorphisms should be regarded as equivalent to the set of local diffeomorphisms and gauge transformations. Moreover, a Lie algebroid isomorphism being a chain map in the exterior algebra sense ensures that isomorphic algebroids are cohomologically equivalent. The Atiyah Lie algebroids derived from principal bundles with common base manifolds and structure groups may therefore be divided into equivalence classes of isomorphic algebroids. Each equivalence class possesses a local representative which we refer to as the trivialized Lie algebroid, and we show that the exterior algebra of the trivialized algebroid gives rise to the BRST complex. We conclude by illustrating the usefulness of Lie algebroid cohomology in computing quantum anomalies, including applications to the chiral and Lorentz-Weyl (LW) anomalies. In particular, we pay close attention to the fact that the geometric intuition afforded by the Lie algebroid (which was absent in the naive BRST complex) provides hints of a deeper picture that simultaneously geometrizes the consistent and covariant forms of the anomaly.  In the algebroid construction, the difference between the consistent and covariant anomalies is simply a different choice of basis. 
\end{abstract}

\newpage

\begingroup
\hypersetup{linkcolor=black}
\tableofcontents
\endgroup

\newpage

\section{Introduction}
\label{sec:intro}
The geometric analysis of gauge theories is a rich area of physics which is deeply interconnected with mathematics \cite{Yang:1954ek,Atiyah:1957,Wu:1975es,Atiyah:1978,RevModPhys.52.175,Jordan:2014uza}. The historical approach to quantifying topological behavior in gauge theories runs through the BRST formalism, which was originally introduced to facilitate the covariant quantization of gauge theories \cite{becchi1974abelian, becchi1976renormalization,Tyutin:1975qk}. It was subsequently realized that the BRST formalism gives rise to an exterior algebra, later dubbed the BRST complex\cite{Bandelloni:1986wz, Henneaux:1989rq, Henneaux:1995ex, Dragon:1996md, DuboisViolette:1991is, Brandt:1996mh, Barnich:2000zw}, which can be used to calculate cohomology classes relevant to quantum anomalies \cite{adler1969axial,bell1969pcac,t1976symmetry,fujikawa1979path,bonora1983some,alvarez1984topological,witten1983global,zumino1985chiral,Nelson:1984gu}. Starting from a principal bundle $P(M,G)$, the basic objective of the BRST complex is to design an exterior algebra that combines the de Rham cohomology of the base manifold $M$ with the cohomology of the local gauge algebra associated with the structure group $G$. The BRST complex accomplishes this task in a series of steps. First, it takes a local section of $P(M,G)$ to define the gauge field $A$, which descends from a bona-fide principal connection. In this way, it forgets about the vertical sub-bundle of $TP$, and restricts its attention only to the de Rham cohomology of the base manifold. Next, the vacuum left behind by the vertical sub-bundle is filled by introducing a graded algebra generated by a set of Grassmann valued fields $c^A(x)$ called ghosts. In this way, one obtains the BRST complex as an exterior bi-algebra consisting of $p$-forms on $M$ contracted with $q$ factors of the ghost field, where the number $q$ is referred to as the ghost number. 

A priori, the ghost fields have no geometric interpretation, rather being interpreted as a computational device. However, it has been argued that a geometric interpretation for the ghost fields exists as the ``vertical components" of an extended gauge field \cite{thierry1980explicit,thierry1980geometrical,quiros1981geometrical,Thierry-Mieg:1985zmg,Neeman:1979cvl, Baulieu:1981sb, ThierryMieg:1982un, ThierryMieg:1987um,Bonora:1981rw,Hirshfeld:1980te, Hoyos:1981pb, Baulieu:1985md,Hull:1990qg,Bonora:2021cxn}. The basic idea behind this interpretation is to contract the ghost fields with the set of Lie algebra generators $c = c^A \otimes \un{t}_A$ and define the extended ``connection" form  $\econn = A + c$ by appending the ghost field to the gauge field. Viewing $\econn$ as a connection, it is natural to define an associated curvature $\ecurv = \td_{\rm BRST}\econn + \frac{1}{2}[\econn,\econn]$, where the coboundary operator of the BRST complex is identified as $\td_{\rm BRST} = \td + \ts$, which is simply the combination of the de Rham differential $\td$ and the BRST operator $\ts$. Enforcing the extra condition that the curvature should have extent only in the de Rham part of the BRST complex, one arrives at a pair of equations defining the action of the BRST operator which can be identified with the Chevalley-Eilenberg differential appearing in Lie algebra cohomology \cite{chevalley1948cohomology,Zumino:1984ws,Brandt:1989gv}. In addition, the action of $\ts$ on the gauge field $A$ can be interpreted as that of an infinitesimal gauge transformation generated by $c(x)$. 

With the ``connection" $\econn$, ``curvature" $\ecurv$, and coboundary operator $\td_{\rm BRST}$ in hand, one can construct ``characteristic classes" in the BRST complex by naively following the Chern-Weil theorem \cite{WeilLetter,zbMATH03070474}. Due to the fact that $\ecurv$ was manufactured to have zero ghost number, the Chern-Simons form associated with a given characteristic class in the BRST complex can be shown to satisfy a series of equations known as the descent equations \cite{Zumino:1984ws,Stora1977,zumino1984chiral,Sorella:1992dr}. One of the resulting equations is the Wess-Zumino consistency condition\cite{wess1971consequences}, which ultimately determines the algebraic form of candidates for quantum anomalies. 

The success of the BRST approach is undeniable. However, it motivates a series of questions. Why should the Grassmann valued fields $c^A(x)$, which started their life in the BRST quantization procedure have an interpretation as the generators of a local gauge transformation? Why is it reasonable to combine the de Rham complex and the ghost algebra into a single exterior bi-algebra? On a related note, why is it reasonable to consider the combination $\econn = A + c$ as a ``connection", and moreover what horizontal distribution does it define? Why should the ``curvature" $\ecurv$ be taken to have ghost number zero, and why does enforcing this constraint turn the BRST operator $\ts$ into the Chevalley-Eilenberg operator for the Lie algebra of the structure group? These are the questions that we will answer in this paper. Quite serendipitously, we will show that there is not an answer to each of these questions individually, but rather each of these individual questions are resolved by the answer to a single question: What is the appropriate geometric interpretation for the BRST complex? Indeed, our main objective will be to demystify the BRST complex once and for all, and in doing so provide a unified geometric picture of quantum anomalies. The mathematical language which is up to this task is that of Lie algebroids \cite{MR0214103, MR0216409, zbMATH01186367, kosmannschwarzbach2002differential, fern2007lie, mackenzie_1987,mackenzie2005general}, the existing uses of which in the context of gauge theories can be found in, e.g., \cite{Ramandi:2014qoa,marle2008differential,lazzarini2012connections, Fournel:2012uv, Carow-Watamura:2016lob, Kotov:2016lpx, Attard:2019pvw,Ciambelli:2021ujl} and the citations therein.

In \cite{Ciambelli:2021ujl} it was argued that the exterior algebra of an Atiyah Lie algebroid derived from a principal $G$-bundle $P(M,G)$ is a geometrization of the physicist's BRST complex. In this note we will provide a novel perspective on this correspondence by elaborating on the concept of the \emph{Lie algebroid trivialization}, which pushes the discussion in \cite{Ciambelli:2021ujl} further. In Section \ref{sec:algebroids} we review the necessary background on Atiyah Lie algebroids, concentrating especially on aspects of the exterior algebra defined therein. In Section \ref{sec:trivialized} we discuss the role of Lie algebroid isomorphisms in facilitating the study of topological aspects of Atiyah Lie algebroids. We introduce an explicit form of Lie algebroid isomorphism between Atiyah Lie algebroids modeled on a commutative diagram, and demonstrate how this isomorphism may be interpreted as implementing both gauge transformations and diffeomorphisms in physical contexts. In Subsection \ref{sec:local} we study Lie algebroid isomorphisms as a tool for trivializing an Atiyah Lie algebroid. We introduce the Lie algebroid atlas which allows for the Lie algebroid trivialization to be carried into the global context. In Subsection \ref{sec:cohomology} we study trivializations of the exterior algebra associated with an Atiyah Lie algebroid, and demonstrate that the resulting cohomology is equivalent to that of the BRST complex. In Section \ref{sec:nonabelian} we apply the lessons from the previous sections to study quantum anomalies. We place an emphasis on the fact that the exterior algebra of the Atiyah Lie algebroid can directly quantify both the consistent and covariant anomaly polynomials. This machinery is applied to the chiral anomaly and the Lorentz-Weyl anomaly in Subsection \ref{sec:examples}. We conclude in Section \ref{sec:disc} in which we provide answers to the questions posed in this introduction, and address directions for follow up work.
\par
This paper is one in a series of ongoing projects intended to synthesize the local properties of gauge theories using the mathematical language of Atiyah Lie algebroids, in route towards a consistent approach to quantizing gauge theories including gravity.

\section{Background on Atiyah Lie Algebroids}
\label{sec:algebroids}
In this section we provide an introduction to Atiyah Lie algebroids focusing on their exterior algebras. We begin by reviewing the construction of an Atiyah Lie algebroid derived from a principal bundle. We subsequently recall the formulation of the exterior algebra of an arbitrary Atiyah Lie algebroid and the coboundary operator $\hatd$. Here, our intention is to include enough detail relevant to the present paper; for more detailed discussions of Lie algebroids, see \cite{Ciambelli:2021ujl} or \cite{mackenzie2005general}.

\subsection{The Lie Algebroid Derived from a Principal Bundle}

Let $P(M,G)$ be a principal $G$-bundle over the base manifold $M$ with structure group $G$. We will denote the Lie algebra of $G$ by $\mg$. The principal bundle $P$ comes equipped with two canonical maps:
\begin{equation}\label{Projection and Right Action}
	\pi: P \rightarrow M\,,\qquad R: P \times G \rightarrow P\,,
\end{equation}
corresponding respectively to the projection and the free right action.

The Atiyah Lie algebroid derived from the principal bundle $P(M,G)$ is given by the vector bundle $A \equiv P \times_{G} TP = TP/G$ over $M$. In particular, $A$ is obtained as the quotient of the tangent bundle $TP$ by the canonically defined right action of $G$. We note that while $TP$ is a bundle over $P$, $A=TP/G$ is importantly a vector bundle over $M$. Furthermore, $A$ is a Lie algebroid because it inherits a bracket algebra from $TP$, denoted by $[\cdot,\cdot]_A$, and possesses an anchor map $\rho$ in the form of the pushforward by the projection, i.e., $\rho = \pi_*: A \rightarrow TM$. Moreover, the map $\rho$ can easily be seen to be surjective, and hence the algebroid $A$ is automatically transitive. This means that we have the following short exact sequence of vector bundles over $M$:
\begin{equation} \label{Short Exact Sequence 1}
\begin{tikzcd}
0
\arrow{r} 
& 
L
\arrow{r}{j} 
&
A
\arrow{r}{\rho} 
& 
TM
\arrow{r} 
&
0 \,.
\end{tikzcd}
\end{equation}
$L$ is the kernel of the anchor map $\rho$, called the isotropy bundle over $M$.
The short exact sequence \eqref{Short Exact Sequence 1} therefore dictates that a section of $A$ can be identified (locally) with the direct sum of a local gauge transformation generated by $\un{\mu} \in \Gamma(L)$ and a diffeomorphism generated by $\un{X} \in \Gamma(TM)$. 

The Atiyah Lie algebroid $A$ has a canonically defined \emph{vertical sub-bundle} $V \subset A$ given by the image of $L$ under the morphism $j$ as $V = j(L)$. This predicates the notion of a Lie algebroid connection as the choice of a \emph{horizontal sub-bundle} which is complimentary to $V$. In the context of the Atiyah Lie algebroid, a connection is quantified by a pair of maps $\omega: A \rightarrow L$ and $\sigma: TM \rightarrow A$ satisfying $\text{ker}(\omega) = \text{im}(\sigma)$, defining a second short exact sequence in the direction opposite to the first one:
\begin{equation} \label{Short Exact Sequence 2}
\begin{tikzcd}
0
\arrow{r} 
& 
L
\arrow{r}{j} 
\arrow[bend left]{l} 
& 
A
\arrow{r}{\rho} 
\arrow[bend left]{l}{\omega}
& 
TM
\arrow{r} 
\arrow[bend left]{l}{\sigma}
&
0\,.
\arrow[bend left]{l} 
\end{tikzcd}
\end{equation}
The map $\omega$ is called the \hlt{connection reform}, and must also satisfy the condition $\omega \circ j = -Id_{L}$. In terms of the connection, the horizontal sub-bundle is given by $H = \text{ker}(\omega) = \text{im}(\sigma)$, and the connection corresponds to a globally defined split of $A$, namely $A=H\oplus V$.

\subsection{The Exterior Algebra of an Atiyah Lie Algebroid}\label{sec:exterior}

The main focus of this work is to analyze the exterior algebra of $A$, denoted by $\Omega(A) = \oplus_{p = 1}^{\text{rank}\,A} \Omega^p(A)$. Each $\Omega^p(A) \equiv \wedge^p A^*$ consists of totally antisymmetric $p$-linear maps from $A^{\otimes p}$ into $C^\infty(M)$. The exterior algebra $\Omega(A)$ has a well-defined coboundary operator $\hatd: \Omega^p(A) \rightarrow \Omega^{p+1}(A)$ determined by the anchor map $\rho$ and the bracket on $A$, via the Koszul formula\cite{chevalley1948cohomology,fernandes2002lie}:
\begin{align} \label{dhat for trivial bundle}
    \hatd\eta(\un{\mX}_1, \ldots, \un{\mX}_{p+1}) ={}& \sum_{i} (-1)^{i+1} \rho(\un{\mX}_i) \eta(\un{\mX}_1, \ldots, \widehat{\un{\mX}_i}, \ldots, \un{\mX}_{p+1})\nn \\
   & + \sum_{i < j} (-1)^{i + j} \eta([\un{\mX}_i, \un{\mX}_j]_A, \un{\mX}_1, \ldots, \widehat{\un{\mX}_i}, \ldots, \widehat{\un{\mX}_j}, \ldots, \un{\mX}_{p+1}) \,,
\end{align}
where $\un{\mX}_1,\ldots,\un{\mX}_{p+1}$ are arbitrary sections on $A$, and $\eta$ a section of $\Omega^p(A)$, with $\eta(\un\mX_1,\ldots,\un\mX_p)\in C^\infty(M)$ the complete contraction of $\eta$ with sections of $A$. 

The exterior algebra $\Omega(A)$ can be extended to $\Omega(A;E)$, namely the exterior algebra on $A$ with values in the vector bundle $E$, by introducing a suitable differentiation of sections of $E$. Such a notion comes in the form of a Lie algebroid representation, which is a morphism $\Aconn{E}: A \rightarrow \Der(E)$ compatible with the anchor. We note that $\Der(E)$ is itself a Lie algebroid, with isotropy bundle given by $\End(E)$ and bracket given via the composition of derivations. The morphism condition simply means that $\Aconn{E}$ has a vanishing curvature:
\begin{equation} \label{Phi E Morphism}
	R^{\Aconn{E}}(\un\mX,\un\mY) := [\Aconn{E}(\un\mX),\Aconn{E}(\un\mY)]_{\Der(E)} - \Aconn{E}([\un\mX,\un\mY]_A) = 0\,,\qquad\forall  \un\mX,\un\mY \in \Gamma(A)\,.
\end{equation}
The compatibility condition ensures that $\Aconn{E}$ maps into a derivation by enforcing the Leibniz-like identity
\begin{equation}
	\Aconn{E}(\un\mX)(f \un\psi) = f \Aconn{E}(\un\mX)(\un\psi) + \rho(\un\mX)(f) \un\psi\,,\qquad \forall  \un\mX \in \Gamma(A)\,,\quad f \in C^{\infty}(M)\,,\quad \un\psi \in \Gamma(E)\,.
\end{equation}
Given such a representation, there is a corresponding Koszul formula generalizing \eqref{dhat for trivial bundle}:
\begin{align} \label{dhat on E}
    \hatd^E\eta(\un{\mX}_1, \ldots, \un{\mX}_{p+1}) ={}& \sum_{i} (-1)^{i+1} \Aconn{E}(\un{\mX}_i) \eta(\un{\mX}_1, \ldots, \widehat{\un{\mX}_i}, \ldots, \un{\mX}_{p+1}) \nn\\
    &+ \sum_{i < j} (-1)^{i + j} \eta([\un{\mX}_i, \un{\mX}_j]_A, \un{\mX}_1, \ldots, \widehat{\un{\mX}_i}, \ldots, \widehat{\un{\mX}_j}, \ldots, \un{\mX}_{p+1}) \,.
\end{align}
The operator $\hatd^E$ can be seen to be nilpotent as a combination of \eqref{Phi E Morphism} and the fact that the bracket on $A$ satisfies the Jacobi identity. For simplicity, we will later refer to the coboundary operator as simply $\hatd$, leaving the particular representation $E$ implicit.

A connection on $A$ specified by $\omega$ and $\sigma$ induces a Lie algebroid representation on any vector bundle $E$ that furnishes a representation space of $L$. Such a representation is determined through the combination of (1) a covariant derivative operator on $E$, $\nabla^E: TM \rightarrow \text{Der}(E)$, and (2) an endomorphism on $E$, $v_E: L \rightarrow \text{End}(E)$. In particular, we take \cite{Ciambelli:2021ujl}
\begin{equation} \label{Basic A Connection}
	\Aconn{E}(\un\mX)(\un\psi) = \nabla^E_{\rho(\un\mX)} \un\psi - v_E \circ \omega(\un\mX) \un\psi\,.
\end{equation}
$\Aconn{E}$ being a Lie algebroid representation through \eqref{Basic A Connection} implies two things. Firstly, $v_E$ must be a morphism, or in other words a linear representation of $L$. Secondly, the curvature of $\nabla^E$ viewed as a connection on $TM$ is determined entirely by the curvature of the horizontal distribution $H$:\cite{Ciambelli:2021ujl}
\begin{equation}\label{Curvature of Nabla}
	R^{\nabla^E}(\un{X},\un{Y}) = [\nabla^E_{\un{X}},\nabla^E_{\un{Y}}]_{\text{Der}(E)} - \nabla^E_{[\un{X},\un{Y}]_{TM}} = -v_E \circ \omega(R^{\sigma}(\un{X},\un{Y}))\,.
\end{equation}

Given the covariant derivative $\nabla^E$, the corresponding connection coefficients are given by
\begin{equation}\label{defSpinConn}
	\nabla^E_{\rho(\un{\mX})} \un{e}_a = {\cal A}^b{}_a(\un{\mX}_H) \un{e}_b\,,
\end{equation}
where $\un{e}_a$ is a basis section of $E$. Hence, we can see that the representation $\Aconn{E}$ acts as
\begin{equation}
	\Aconn{E}(\un\mX)(\un{e}_a) = \Big({\cal A}^b{}_a(\un{\mX}_H)  - (v_E(\omega(\un\mX_V)))^b{}_a\Big) \un{e}_b\,.
\end{equation}

\section{Lie Algebroid Isomorphisms}
\label{sec:trivialized}

Given that $\hatd$ is nilpotent on $\Omega(A,E)$, it provides a well-defined notion of cohomology, which we refer to as \hlt{Lie algebroid cohomology}. In this section, our intention is to explain how this cohomology is related to the usual notion of BRST cohomology. In \cite{Ciambelli:2021ujl}, it was shown that the action of $\hatd$ can be thought of as containing within it the BRST transformation. In this section, we will emphasize the role played by isomorphisms of Lie algebroids. We will show that two Lie algebroids with connection that are related by an isomorphism are different representatives of a topological class, and the cohomology of the respective $\hatd$ agree. In this sense, the $\hatd$ cohomology is invariant under isomorphism. In \cite{Ciambelli:2021ujl} the notion of a local trivialization of a Lie algebroid was reviewed. This is a map $\tau_U:A\big|_U\to TU\oplus L\big|_U$, with $U$ an open subset of the base manifold $M$, through which the connection on $A$ can be expressed locally as a gauge field. We will show below that it is in this description that the usual physics notation $\hatd_{\tau} \to \td + s$ makes sense. This isomorphism may then be used to relate Lie algebroid cohomology to the usual physics notions of BRST cohomology. 

\subsection{A Commutative Diagram}\label{sec:morphisms}

A Lie algebroid morphism is a map $\varphi: A_1 \rightarrow A_2$ between two Lie algebroids, which preserves the geometric structure of the Lie algebroids as encoded in their brackets. That is, for all $\un\mX,\un\mY \in \Gamma(A_1)$,
\begin{equation} \label{Morphism Condition}
    R^\varphi(\un\mX,\un\mY):=-\varphi([\un\mX , \un\mY]_{A_1}) + [\varphi(\un\mX), \varphi(\un\mY)]_{A_2}=0\,.
\end{equation}
In this section we focus on a subclass of Lie algebroid morphisms which are, in fact, isomorphisms of the underlying vector bundles. Consider a set of Lie algebroids that share the same base manifold and structure group. In general, two such algebroids may be topologically distinct. Our goal is to emphasize that two algebroids in this set, $A_1$ and $A_2$, will be topologically equivalent if there exists an isomorphism between them. To accomplish this goal, we seek to understand the conditions under which the set of structure maps of two Lie algebroids define a commutative diagram of the following form:
\begin{equation}
\label{Transitive Lie Algebroid Morphism}
\begin{tikzcd}
& 
& 
A_1
\arrow[left]{dd}{\varphi}
\arrow[left]{dl}{\omega_1}
\arrow[bend left]{dr}{\rho_1}
& 
&
\\
0
\arrow{r}
&
L
\arrow[bend left]{l} 
\arrow[bend left]{ur}{j_1}
\arrow[bend right, swap]{dr}{ j_2}
&
&
TM
\arrow{r}
\arrow[shift left]{ul}{\sigma_1} 
\arrow[swap]{dl}{\sigma_2} 
&
0\,.
\arrow[bend left]{l} 
\\
& 
&
A_2
\arrow[left, swap]{ul}{\omega_2}
\arrow[bend right,swap]{ur}{\rho_2}
\arrow[shift left]{uu}{\overline{\varphi}}
&
&
\end{tikzcd}
\end{equation}
Note that $J\equiv\sigma_2 \circ \rho_1$ is a map from $H_1$ to $H_2$, while $K \equiv j_2 \circ \omega_1$ is a map from $V_1$ to $V_2$. Clearly, we can write $\varphi=J-K$. Our motivation for considering \eqref{Transitive Lie Algebroid Morphism} is that it respects the horizontal and vertical splittings of the two algebroids, and will subsequently provide a useful physical picture for general Lie algebroid isomorphisms.\footnote{Here, we are discussing isomorphisms using an {\it active} language; in the corresponding {\it passive} description, an isomorphism would be understood as a change of basis for the same algebroid.}

By commutativity, the maps $\varphi$ and $\overline{\varphi}$ in \eqref{Transitive Lie Algebroid Morphism} apparently define isomorphisms of the vector bundles $A_1$ and $A_2$. However, it is not immediately clear that these maps respect the algebras defined by the brackets on these bundles. To this end, we will now demonstrate that the map $\varphi$ will be a Lie algebroid morphism if and only if the horizontal distributions of $A_1$ and $A_2$ as defined by their respective connections $\omega_1$ and $\omega_2$ share the same curvature. Recall that the curvature of a connection reform $\omega$ is the horizontal $L$-valued form given by\footnote{We have introduced the graded Lie bracket between $L$-valued differential forms. For $\alpha \in \Omega^m(A; L)$ and $\beta \in \Omega^n(A; L)$, $[\alpha,\beta]_{L}$ is defined as
\begin{equation*}
	[\alpha,\beta]_{L}(\un\mX_1, \ldots, \un\mX_{m+n}) = \sum_{\sigma }\text{sgn}(\sigma) [\alpha(\un\mX_{\sigma(1)}, \ldots, \un\mX_{\sigma(m)}), \beta(\un\mX_{\sigma(m+1)}, \ldots, \un\mX_{\sigma(m+n)})]_{L}\,,
\end{equation*}
where $\un\mX_1,\ldots ,\un\mX_{m+n}$ are arbitrary sections on $A$, $\sigma$ denotes the permutations of $(1,\ldots,m+n)$, and $\text{sgn}(\sigma)=1$ for even permutations and  $\text{sgn}(\sigma)=-1$ for odd permutations.} 
\beq
\Omega = \hatd\omega + \frac{1}{2}[\omega,\omega]_L\,.
\eeq
Suppose the curvatures of $\omega_1$ and $\omega_2$ are $\Omega_1$ and $\Omega_2$, respectively. We can compute that
\begin{align}
R^{\varphi}(\un\mX_H,\un\mY_H)&= R^{\sigma_2}(\rho_1(\un\mX_H),\rho_1(\un\mY_H))+j_2(R^{-\omega_1}(\un\mX_H,\un\mY_H)\nn\\
&= j_2(\Omega_2(\varphi(\un\mX),\varphi(\un\mY)))-j_2(\Omega_1(\un\mX,\un\mY))\,,
\end{align}
where we used $\varphi=J-K$ and Eq.~(43) of \cite{Ciambelli:2021ujl}:
\beqn\label{Russianformula}
R^\sigma(\rho(\un\mX),\rho(\un\mY))=j(\Omega(\un\mX,\un\mY))=-j(R^{-\omega}(\un\mX_H,\un\mY_H))\,.
\eeqn 
In this way, we see that $\varphi$ will be a morphism of the brackets if and only if
\beq\label{Curvature pullback}
\Omega_1(\un\mX,\un\mY) =\Omega_2(\varphi(\un\mX),\varphi(\un\mY))\,.
\eeq

Provided $\varphi$ is an isomorphism, it will induce a linear transformation on bundles associated to $A_1$ and $A_2$ to preserve Lie algebroid representations. Let $E_1$ and $E_2$ be isomorphic vector bundles over $M$ which are associated, respectively, to $A_1$ and $A_2$ by Lie algebroid representations $\Aconn{E_j}: A_j\to\text{Der}(E_j)$, with $j=1,2$. 
Then, accompanying the Lie algebroid isomorphism $\varphi$, there is a corresponding map on the associated bundles, which can be written as
\begin{equation}
    g_\varphi: E_1 \rightarrow E_2\,.
\end{equation}
By construction, we enforce that this map is compatible with the Lie algebroid representations of $A_1$ and $A_2$ in the sense that
\beq \label{Iso-Rep compatibility}
	\Aconn{E_2} \circ \varphi(\un\mX)(g_{\varphi}(\un{\psi})) = g_{\varphi}(\Aconn{E_1}(\un\mX)(\un{\psi}))\,, \qquad \forall \un\mX \in \Gamma(A_1)\,,\quad \un{\psi} \in \Gamma(E_1)\,.
\eeq
Let $\varphi^*: \Omega(A_2;E_2) \rightarrow \Omega(A_1;E_1)$ denote the Lie algebroid pullback map induced by $\varphi$. Explicitly, given $\eta \in \Omega^r(A_2;E_2)$ and $\un\mX_1, \ldots, \un\mX_r \in \Gamma(A_1)$ we have
\begin{equation}\label{Lie algebroid pullback}
	(\varphi^*\eta)(\un\mX_1, \ldots, \un\mX_r) =g_\varphi^{-1}\big(\eta(\varphi(\un\mX_1), \dots, \varphi(\un\mX_r))\big)\,. 
\end{equation}
Using this notation along with the properties \eqref{Morphism Condition} and \eqref{Iso-Rep compatibility} it is easy to establish that 
\beq\label{Chain Map condition}
\hatd_1\circ \varphi^*=\varphi^*\circ\hatd_2\,,
\eeq
which means that $\varphi$ is a \hlt{Lie algebroid chain map} in the exterior algebra sense. For an explicit demonstration, we refer the reader to Appendix \ref{app:chainmap}. 

Using \eqref{Lie algebroid pullback} we can rewrite \eqref{Curvature pullback} as
\beq \label{Curvature pullback 2}
	\Omega_1 = \varphi^*\Omega_2\,.
\eeq
Eq.~\eqref{Curvature pullback 2} indicates that a Lie algebroid isomorphism of the form \eqref{Transitive Lie Algebroid Morphism} involves a topological consideration about the algebroids in question. In Section \ref{sec:nonabelian} we introduce a version of the Chern-Weil homomorphism which is applicable to Lie algebroid cohomology. This will provide a recipe for constructing Atiyah Lie algebroid cohomology classes in terms of characteristic polynomials in curvature. Recall that a characteristic class satisfies a so-called ``naturality" condition, which essentially implies that the pullback commutes through the characteristic class; i.e., if $\lambda(\Omega)$ is a characteristic class of a curvature $\Omega$, then 
\beq
\lambda(\varphi^*\Omega) = \varphi^*\lambda(\Omega)\,.
\eeq 
Hence, two Lie algebroids whose curvatures are related as \eqref{Curvature pullback} will possess an isomorphism between their cohomologies. Eq.~\eqref{Chain Map condition} similarly implies that isomorphic Lie algebroids possess isomorphic cohomology classes. In light of these observations, we can view the Lie algebroid isomorphism as a device for organizing the set of Atiyah Lie algebroids with connection into topological equivalence classes. Let $(A,\omega)$ denote an Atiyah Lie algebroid $A$ with connection reform $\omega$. Then,
\begin{equation}
[(A,\omega)] := \{(A',\omega') \; | \; \exists \varphi: A \rightarrow A' \text{ s.t. } \Omega = \varphi^*\Omega' \}
\end{equation}
can be regarded as the set of topologically equivalent Atiyah Lie algebroids with connection. 

From a physical perspective Eqs.~\eqref{Curvature pullback} and \eqref{Chain Map condition} establish the fact that the commutative diagram \eqref{Transitive Lie Algebroid Morphism} encodes diffeomorphisms and gauge transformations relating isomorphic Lie algebroids. In particular, it is straightforward to establish that the connection coefficients [see Eq.~\eqref{defSpinConn}] satisfy
\begin{align}
\label{Gauge Transformation}
({\cal A}_1)_{\un\alpha_1}{}^{a_1}{}_{b_1}&=J^{\un\alpha_2}{}_{\un\alpha_1}(g_\varphi^{-1})^{a_1}{}_{a_2}\Big(({\cal A}_2)_{\un\alpha_2}{}^{a_2}{}_{b_2}+\delta^{a_2}{}_{b_2}\rho(\un E_{\un\alpha_2})\Big)g_\varphi^{b_2}{}_{b_1}\,,\\
(v_E(\omega_1))_{\un A_1}{}^{a_1}{}_{b_1}&=K^{\un B_2}{}_{\un A_1}(g_\varphi^{-1})^{a_1}{}_{a_2}(v_E(\omega_2))_{\un B_2}{}^{a_2}{}_{b_2}g_\varphi^{b_2}{}_{b_1}\,.
\end{align}
That is, the components of ${\cal A}$ and $\omega$ transform like a gauge field and a gauge ghost, respectively. 
Eq.~\eqref{Gauge Transformation} is compatible with \eqref{Curvature pullback}; recall that the curvatures of gauge fields related by a gauge transformation are equivalent up to a conjugation. In this respect, we can also identify the Lie algebroid isomorphism \eqref{Transitive Lie Algebroid Morphism} as encoding the data of a gauge transformation. In other words, the set $[(A,\omega)]$ can be regarded as an orbit of gauge equivalent algebroids. In a separate work \cite{klinger2023abc}, we use this remark to construct the \hlt{configuration algebroid}, which can be regarded as a concise definition of the space of gauge orbits of connections that can be employed in any gauge theory formulated in terms of Atiyah Lie algebroids. 

\subsection{Local Trivialization of an Atiyah Lie Algebroid}
\label{sec:local}

In the last section we have  shown that there exists a Lie algebroid isomorphism of the form \eqref{Transitive Lie Algebroid Morphism} between Lie algebroids with connection whose horizontal distributions have curvatures related by \eqref{Curvature pullback}. It is perhaps worth mentioning that this very same construction was used in constructing a representation of a Lie algebroid $A$ by the Lie algebroid $\Der(E)$, for some associated vector bundle $E$. In fact, this is a slight generalization of what we presented above, in that whereas the isomorphism in question is $\phi_E:A\to \Der(E)$, these two algebroids do not share the same isotropy bundle, but instead there is a further isomorphism $v_E:L\to \End(E)$ between them. Locally this isomorphism can be thought to give a matrix representation (on the fibres of $E$) of the Lie algebra. 

A local trivialization of a Lie algebroid can also be thought of as an example of a Lie algebroid isomorphism, with the details presented in terms of local data. Using the notation of \cite{Ciambelli:2021ujl}, on open sets $U_i\subset M$, we have\footnote{Here $A^{U_i}$ is the restriction of the Lie algebroid $A$ to the local neighborhood $U_i \subset M$. In other words, $A^{U_i}$ is a vector bundle over $U_i$.}
\beq\label{localtriv}
\tau_i:A^{U_i}\to TU_i\oplus L^{U_i}\,,
\eeq
and so local sections of $A$ can be expressed in terms of local bases for $TM$ and $L$
\beq
\tau_i(\un\mX_H)=\mX_{i,H}^\alpha\tau_i{}^\mu{}_\alpha(\un\pa^{U_i}_\mu+b_{i}{}_\mu^A\un t^{U_i}_A)\,,\qquad
\tau_i(\un\mX_V)=\mX_{i,V}^{\un A}\tau_i{}^A{}_{\un A}\un t^{U_i}_A\,.
\label{deftau}
\eeq
The coefficients $b_{i}{}_\mu^A$ are the components of a $\mg$-valued 1-form on $M$, that transforms on overlapping open sets as a gauge field by consequence of \eqref{Gauge Transformation}.
\begin{figure}[!htbp]
\center
\includegraphics[width=1.7in]{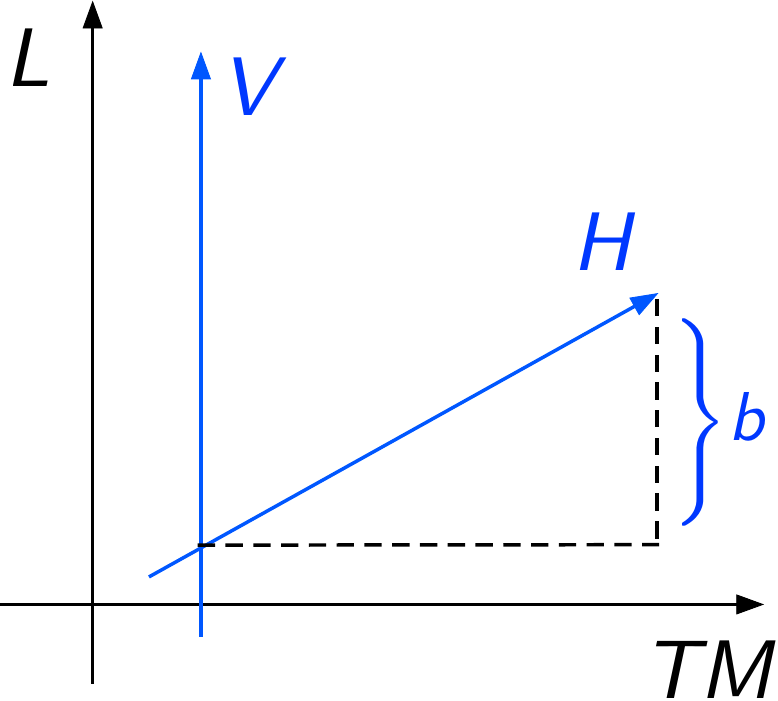}
\caption{A visualization of a Lie algebroid. A connection gives a global split $A=H\oplus V$, which locally can be viewed as determined by a gauge field $b$ defined with respect to ``axes'' corresponding to sub-bundles $TM$ and $L$.}
\label{fig:VHcartoon.pdf}
\end{figure}

As we have now established, in each open set $U_i \subset M$, we realize a Lie algebroid isomorphism $\tau_i: A^{U_i} \rightarrow TU_i \oplus L^{U_i}$.\footnote{Note that here we are using the notion of isomorphism in the active sense, and hence we distinguish $A^{U_i}$ from $TU_i \oplus L^{U_i}$. In what follows, the reader may find it profitable to think from a passive perspective: indeed our use of $A^{U_i}$ versus $TU_i \oplus L^{U_i}$ can be thought of as simply corresponding to a different choice of basis, the first natural from the $H\oplus V$ split, the second natural from the local $TU\oplus L$ split.} Suppose $\{U_i\}$ form an open cover for the base $M$, then we can sew together the aforementioned local charts to obtain a Lie algebroid atlas. Sewing the charts $\tau_i$ together requires that we also specify transition functions $t_{ij}: A^{U_i} \rightarrow A^{U_j}$, which are Lie algebroid isomorphisms with support in the intersection $U_i \cap U_j$ for each pair of $U_i$ and $U_j$. This is equivalent to the perhaps more familiar notion that overlapping charts in a principal bundle must agree up to a gauge transformation. The presence of non-trivial transition functions in the algebroid context ensures that topological data is preserved under trivialization. Together, the collection $\{U_i, \tau_i, t_{ij}\}$ carries the intuition of the Lie algebroid trivialization into a global context. In the following we will use the abbreviated notation $\tau: A \rightarrow A_{\tau}$ to refer to the local Lie algebroid isomorphism mapping $A$ into the trivialized Lie algebroid $A_{\tau} \simeq TU \oplus L^{U}$ for some $U \subset M$. That is, the notation $A_{\tau}$ serves to remind that $A_{\tau}$ involves restricting $A$ to an open set. We leave the open subset $U$ unspecified with the understanding that the Lie algebroid atlas allows for the algebroid $A$ to be trivialized when restricted to any open neighborhood of the base.

To be precise about details, we will introduce explicit bases for the various vector bundles; although we will not indicate so, these should be understood to be valid locally on some open set of $M$. So we introduce the notation for bases for the bundles $TM$ and $L$ and their dual bundles:
\begin{equation} \label{Coordinate Bases}
\begin{split}
	TM &= \text{span}\{\un{\partial}_{\mu}\}\,,\qquad T^*M = \text{span}\{\td x^{\mu}\}\,,\qquad \mu=1,\ldots, \dim M\,,\\
	L &= \text{span}\{\un{t}_A\}\,,\qquad L^* = \text{span}\{t^A\}\,,\qquad A=1,\ldots, \dim G\,.
\end{split}
\end{equation}
These bases are dual in the sense that
\begin{equation}
	\td x^{\mu}(\un{\partial}_{\nu}) = \delta^{\mu}{}_{\nu},\qquad t^A(\un{t}_B) = \delta^A{}_B\,,\qquad\td x^{\mu}(\un{t}_A) = 0\,,\qquad t^A(\un{\partial}_{\mu}) = 0\,.
\end{equation}
Given the above notation, we have a choice to make for a basis of sections of the trivialized Lie algebroid $A_\tau$. We will refer to such choices as  ``splittings", and we will make reference to two natural choices which we refer to as the \emph{consistent splitting} and the \emph{covariant splitting}, respectively. The relevance of this nomenclature will become clear shortly. These two splittings correspond in fact to the two sets of axes shown in Figure~\ref{fig:VHcartoon.pdf}, and they are distinguished precisely because of the non-trivial connection on $(A_\tau,\omega_\tau)$.

By a covariant splitting, we mean a split basis as described in \cite{Ciambelli:2021ujl}. Consider an algebroid $(A,\omega)$ for which we take a basis of sections $\{\un E_{\un\alpha},\un E_{\un A}\}$ (where $\un\alpha=1,\ldots,\dim M$, $\un A=1,\ldots,\dim G$). Such a basis has the virtue that $\omega(\un E_{\un\alpha})=\un 0$ and $\rho(\un E_{\un A})=\un 0$, namely they span $H$ and $V$ respectively. Given the map $\tau$, it is natural to choose a basis $\{\tau(\un E_{\un\alpha}),\tau(\un E_{\un A})\}$ for $A_\tau$. Since we will now deal directly with $A_\tau$, we will for brevity denote such a basis by $\{\un {\hat E}_{\un\alpha},\un {\hat E}_{\un A}\}$. Thus  a covariant splitting corresponds to a choice of basis sections that are aligned with the global split $A_\tau=H_\tau\oplus V_\tau$. Locally, these sections can be expressed in terms of the bases for $TM$ and $L$ as
\beqn
\un {\hat E}_{\un\alpha}=\rho_\tau^\mu{}_{\un\alpha}(\un\pa_\mu+b_\mu^A\un t_A)\,,\qquad 
\un {\hat E}_{\un A}=-\omega_\tau^A{}_{\un{A}}\un t_A\,,
\eeqn
while the dual bases can be written as (also referred to as the ``mixed local basis'' in \cite{Fournel:2012uv})
\beq\label{taudualbasis}
\hat E^{\un\alpha}=\sigma_\tau^{\un{\alpha}}{}_{\mu} \td x^\mu\,,\qquad
{\hat E}^{\un A}=j_\tau^{\un{A}}{}_{A}(t^A-b^A_\mu \td x^\mu)\,.
\eeq
Suppose $\un X=X^\mu\un\pa_\mu\in \Gamma(TM)$ and $\un\mu=\mu^A\un t_A\in\Gamma(L)$, then we have a section $\un\mX$ of $A_\tau$ which can be expressed in this covariant splitting as
\beq\label{covariantsection}
\un\mX=X^\mu \sigma_\tau(\un\pa_\mu)-\mu^A j_\tau(\un t_A)=X^\mu \sigma_\tau^{\un\alpha}{}_\mu\un {\hat E}_{\un\alpha}-\mu^A j_\tau^{\un A}{}_A \un {\hat E}_{\un A}=X^\mu(\un\pa_\mu+b_\mu^A\un t_A)+\mu^A\un t_A\,.
\eeq

On the other hand, by a consistent splitting, we mean a choice of basis for $A_\tau$ that is aligned with the bases for $TM$ and $L$. That is, in the consistent splitting, we can write a section of $A_\tau$ as
\beq
\un\mX=\mX^\mu\un\pa_\mu+\mX^A\un t_A\,.
\eeq
By comparing to the covariant split \eqref{covariantsection}, we see that 
\beq
\mX^\mu=X^\mu\,,\qquad \mX^A=\mu^A+X^\mu b_\mu^A\,,
\eeq
and thus in the consistent splitting, the gauge field is contained in an off-block-diagonal piece of $\sigma_\tau$.

In the current set up, the connection reform $\omega_\tau$ which defines the horizontal distribution through its kernel can be written in the consistent splitting as
\begin{equation} \label{omega = b - c}
	\omega_\tau = \omega_\tau^A{}_{\un{A}} \hat E^{\un{A}} \otimes \un{t}_A = \omega_\tau^A{}_{\un{A}}j_\tau^{\un{A}}{}_B (t^B -  b^B_{\mu} \td x^{\mu}) \otimes \un{t}_A = (b^A_{\mu} \td x^{\mu} - t^A) \otimes \un{t}_A = b - \varpi\,.
\end{equation}
where we defined
\beq \label{MC on L}
\varpi = \varpi^A\otimes\un{t}_A=t^A \otimes \un{t}_A\,,
\eeq
 which can be interpreted as the Maurer-Cartan form on $L$. Recall that $L$ is a bundle of Lie algebras, which means that the $\varpi$ given in \eqref{MC on L} should be interpreted as the Maurer-Cartan form for the group $G$ pointwise on the base manifold $M$. In other words, $\varpi$ is a field of Maurer-Cartan forms, with $\varpi(x)$ being the Maurer-Cartan form for each fiber of $L$ at $x \in M$. The spatial dependence of $\varpi$ will play a significant role in defining the exterior algebra in the consistent splitting.
 
Eq.~\eqref{omega = b - c} explicitly shows that the connection reform can be understood as the sum of two pieces, the first related to the gauge field, and the second related to the Maurer-Cartan form of the gauge algebra, if we interpret it in the consistent splitting (i.e., in terms of the bases for $TM$ and $L$ and their duals). This equation should be compared with the idea of an extended ``connection" in the BRST complex which is typically taken to be of the form $\econn = A + c$ where $A$ is a local gauge field and $c$ is the ghost field \cite{thierry1980explicit,thierry1980geometrical,
quiros1981geometrical}. However, Eq.~\eqref{omega = b - c} has an advantage over the conventional extended ``connection" because it possesses a manifestly geometric interpretation as a genuine connection in the algebroid context. 

\subsection{The Cohomology of Trivialized Lie Algebroids}
\label{sec:cohomology}
We now turn our attention to the main focus of this section---understanding the exterior algebra of the trivialized algebroid. 
The bracket on $A_{\tau}$ can be written explicitly for the basis sections as
\begin{align} \label{HH}
[\un{\hat E}_{\un{\alpha}}, \un{\hat E}_{\un{\beta}}]_{A_{\tau}}& = \sigma_\tau\left([\rho_\tau(\un{\hat E}_{\un{\alpha}}), \rho_\tau(\un{\hat E}_{\un{\beta}})]_{TM}\right)
+j_\tau( \Omega_{\un{\alpha} \un{\beta}})\,,
\\
\label{HV}
[\un{\hat E}_{\un{\alpha}}, \un{\hat E}_{\un{B}}]_{A_\tau}& = 
-j_\tau\left(R^{-\omega_\tau}(\un{\hat E}_{\un\alpha},\un{\hat E}_{\un B})\right)
=j_\tau\left(\nabla^L_{\un{\hat E}_{\un\alpha}}(\omega_\tau^A{}_{\un B}\un t_A)\right)=j_\tau\left(\Aconn{L}(\un{\hat E}_{\un\alpha})(\omega_\tau^A{}_{\un B}\un t_A)\right)\,,
    \\
\label{VV}
[\un{\hat E}_{\un{A}}, \un{\hat E}_{\un{B}}]_{A_\tau} &= j_\tau\left([\omega_\tau(\un{\hat E}_{\un{A}}),\omega_\tau(\un{\hat E}_{\un{B}})]_L\right)
=-\omega_\tau^A{}_{\un A}\omega_\tau^B{}_{\un B}f_{AB}{}^C\un{\hat E}_{\un C}j_\tau^{\un C}{}_C\,.
\end{align}

The coboundary operator for the complex $\Omega(A_{\tau}; E)$, denoted by $\hatd_{\tau}$, is defined precisely by the Koszul formula \eqref{dhat on E}. In terms of the isomorphism $\tau:A\to A_\tau$, we have, as in \eqref{Chain Map condition}, $\hatd\circ\tau^*=\tau^*\circ\hatd_\tau$. Working in $A_{\tau}$, we now have two different ways of splitting $\Omega(A_{\tau};E)$ into a bi-complex. Firstly, we can use the covariant splitting of $A_{\tau}$ to identify
\begin{equation}
	\Omega^p(A_{\tau};E) = \bigoplus_{r + s = p} \Omega^{(r,s)}(H_{\tau},V_{\tau};E)\,,
\end{equation}
where $\Omega^{(r,s)}(H_{\tau},V_{\tau};E)$ consists of bi-forms of degree $r$ in the algebra of $H_{\tau}$ and degree $s$ in the algebra of $V_{\tau}$. This is certainly the most natural splitting of the exterior algebra, as it is globally defined given a connection. We will show that this is equivalent to, but not the same as, the usual splitting, where $r$ counts the de Rham form degree and $s$ counts ghost number.

Alternatively, using the consistent splitting for $A_{\tau}$ we can identify
\begin{equation}
	\Omega^p(A_{\tau};E) = \bigoplus_{r + s = p} \Omega^{(r,s)}(TM,L;E)\,,
\end{equation}
where $\Omega^p(A_{\tau};E)$ now consists of bi-forms of degree $r$ in the de Rham cohomology of $M$ and degree $s$ in the Chevalley-Eilenberg algebra of $L$.

To understand precisely how this works, we consider the action of $\hatd_{\tau}$ on sections of various bundles. We will show that the action of $\hatd_\tau$ can be interpreted as acting as $\td+\ts$ on the components of sections, reproducing the usual physics notation \cite{Ciambelli:2021ujl} (apart from the fact that the usual Grassmann quantities appear instead as forms).

As a  first example, we consider an $E$-valued scalar $\un{\psi} = \psi^a \un{e}_a\in\Gamma(E)$. Using the Koszul formula, we have
\begin{align}
\hatd_{\tau} \un{\psi}
&=\hat E^{\un M}\otimes\phi_E(\un {\hat E}_{\un M})(\un\psi)\nn\\
&=\rho_\tau^{\mu}{}_{\un{\alpha}}\left(\partial_{\mu} \psi^a + v_E(b_{\mu})^a{}_b \psi^b\right)   \hat E^{\un{\alpha}}\otimes \un{e}_a- v_E(\omega_{\hat{A}})^a{}_b \psi^b \;   E^{\hat{A}}\otimes\un{e}_{a}\nn\\
&=
 \Big(\td \psi^a
 +  v_E(\un t_A)^a{}_b\varpi^A \psi^b  \Big)\otimes\un{e}_a \,,
\end{align}
which we identify with\footnote{It should be noted that in \cite{Ciambelli:2021ujl} this was written as $\hatd\un\psi=\nabla^E\un\psi+\ts\un\psi$. These results are consistent, given that $\hatd\un\psi=\nabla^E\un\psi+\psi^a \ts\un e_a+ \ts\psi^a\otimes\un e_a=\td\psi^a\otimes \un e_a+ \ts\psi^a\otimes\un e_a$. This is a general feature: by extracting the basis elements, the gauge fields in the covariant derivative are canceled by those coming from $\ts\un e_a$. We will see this pattern repeated in additional examples.}
\beq\label{hatddplusE}
\hatd_{\tau} \un{\psi}=  (\td+\ts)\psi^a\otimes\un{e}_a\,,
\eeq
if we interpret 
\beq 
\ts\psi^a:= v_E(\un t_A)^a{}_b\varpi^A \psi^b\,.
\eeq 

As a second example, consider a section $\beta\in\Gamma(A_\tau^*\times E)$. Employing the Koszul formula (which is most easily employed by translating $\beta$ into the covariant split basis), we find 
\begin{align}
\hatd_\tau\beta={}&\frac12\hat E^{\un M}\wedge \hat E^{\un N}\otimes\Big(
\phi_E(\un {\hat E}_{\un M})(\beta^a_{\un N}\un e_a)
-\phi_E(\un {\hat E}_{\un N})(\beta^a_{\un M}\un e_a)
-\beta([\un {\hat E}_{\un M},\un {\hat E}_{\un N}]_{A_\tau})\Big)\nn\\
={}&\Big[\Big(\td (\sigma_\tau^{\un\alpha}{}_\nu\beta^a_{\un\alpha}-j_\tau^{\un B}{}_B\beta^a_{\un B}b_\nu^B) +  v_E(\un t_A)^a{}_bt^A (\sigma_\tau^{\un\alpha}{}_\nu\beta^a_{\un\alpha}-j_\tau^{\un B}{}_B\beta^a_{\un B}b_\nu^B)  \Big)\wedge \td x^\nu
\nn\\&
+\Big(\td (j_\tau^{\un B}{}_B\beta^a_{\un B}) +  v_E(\un t_A)^a{}_bt^A (j_\tau^{\un B}{}_B\beta^b_{\un B}) -\frac12f_{AB}{}^{C}(j_\tau^{\un B}{}_C\beta^b_{\un B}) t^A
 \Big)\wedge t^B
\Big]\otimes\un e_a\,.
\end{align}
Recognizing $\beta^a_\nu=\sigma_\tau^{\un\alpha}{}_\nu\beta^a_{\un\alpha}-j_\tau^{\un B}{}_B\beta^a_{\un B}b_\nu^B$ and $\beta^a_A=j_\tau^{\un B}{}_A\beta^a_{\un B}$, we have
\beq
\hatd_\tau\beta=
\Big(\td \beta^a_\nu +  v_E(\un t_A)^a{}_bt^A \beta^a_\nu\Big)\wedge \td x^\nu\otimes\un e_a
+\Big(\td \beta^a_B +  v_E(\un t_A)^a{}_bt^A \beta^b_B -\frac12f_{AB}{}^{C}\beta^a_C t^A
 \Big)\wedge t^B\otimes\un e_a\,,
\eeq
so we see that
\beq
\hatd_\tau\beta=(\td+\ts)\beta^a_\mu\wedge \td x^\mu\otimes\un e_a + (\td+\ts)\beta^a_A\wedge t^A\otimes\un e_a\,,
\eeq
if
\beq
\ts\beta^a_\nu=v_E(\un t_A)^a{}_b\varpi^A \beta^a_\nu\,,\qquad
\ts\beta^a_B=v_E(\un t_A)^a{}_b\varpi^A \beta^b_B -\frac12f_{AB}{}^{C}\beta^b_C \varpi^A\,.
\eeq
We note that this is of a similar form to the previous example in \eqref{hatddplusE}.

As a final example, we consider the connection reform $\omega_\tau$, which we regard as an element of $\Omega^1(A_\tau,L)$. We have
\begin{align}
\hatd_\tau\omega_\tau&= \hatd_\tau(b-\varpi)\nn\\
&=(\Omega_\tau^A-\frac12 f_{BC}{}^A\omega_\tau^B\wedge \omega_\tau^C)\otimes\un t_A
\label{defOmega}\\
&= (\td b^A+f_{BC}{}^A\varpi^B\wedge b^C-\frac12 f_{BC}{}^A \varpi^B\wedge\varpi^C)\otimes\un t_A\,,\label{hatdtauomega}
\end{align}
where in the last line we made use of the result \eqref{omega = b - c}, writing $\varpi=\varpi^A\otimes\un t_A$.

We note that if we identify
\beq \label{s of omega and varpi}
\ts b^A=\td\varpi^A+f_{BC}{}^A\varpi^B\wedge b^C\,,\qquad \ts\varpi^A=\frac12 f_{BC}{}^A \varpi^B\wedge\varpi^C\,,
\eeq
then we obtain
\beq\label{hatdomegads}
\hatd_\tau\omega_\tau=(\td+\ts)\omega_\tau^A\otimes\un t_A \,.
\eeq
To understand \eqref{hatdomegads} one must establish an interpretation for the $\td\varpi^A$ in \eqref{s of omega and varpi}. As we have alluded to below \eqref{MC on L}, $\varpi$ is not spatially constant, and therefore has a nonzero derivative under de Rham $\td$. Considering the following pair of facts: 
\beq
	\hat{i}_{-j(\un\mu)}\varpi^A=-\mu^A\,, \qquad \hat{\cal L}_{-j(\un\mu)}\varpi^A=0\,,\qquad \forall \un\mu\in\Gamma(L)\,,
\eeq
and noticing that $\hat{\cal L}_{\un\mX}=\hat{i}_{\un\mX}\hatd+\hatd\hat{i}_{\un\mX}$, we have
\beq \label{de Rham of varpi}
	\hat{i}_{-j(\un\mu)} \td \varpi^A = \td \mu^A\,.
\eeq
Then, the first equation in \eqref{s of omega and varpi} is consistent with the standard variation of the gauge field:
\beq
	\hat{i}_{-j(\un\mu)}\ts b^A=\td\mu^A+[b,\un\mu]^A\,.
\eeq
Therefore, starting from the formal definition \eqref{dhat on E} of the nilpotent coboundary operator in the algebroid exterior algebra, we established the relationship  between $\hatd_{\tau}$ and the BRST differentiation $\ts$. Again, we emphasize that this result is a natural consequence of the geometric structure of the algebroid. 

\section{Anomalies from Lie Algebroid Cohomology}
\label{sec:nonabelian}

We have now demonstrated that the fundamental features of the BRST complex are geometrically encoded in the Atiyah Lie algebroid. Working in the consistent splitting, the exterior algebra of the trivialized algebroid is a bi-complex consisting of differential forms on the base manifold $M$ and differential forms in the exterior algebra associated to the local gauge group. This is the state of affairs described in the BRST complex but only after making a series of choices \cite{becchi1976renormalization,alvarez1984topological,zumino1985chiral,
zumino1984chiral,kanno1989weil}. We have now shown why these choices are reasonable. For example, the counterpart of the extended  ``connection" $\econn = A + c$ is identified with $\omega_\tau = b - \varpi$ in the algebroid context; $b$ corresponds to the gauge field $A$, and $\varpi$ corresponds to the ghost field $c$ (up to a sign difference). Significantly, $\omega_\tau$ is a genuine connection which defines a horizontal distribution on the algebroid. Moreover, the appearance of the Maurer-Cartan form $\varpi$ justifies the interpretation of the ghost field $c$ in the BRST formalism as a generator of gauge transformations. 

As discussed in \cite{Ciambelli:2021ujl}, the ``Russian formula" central to the BRST analysis (see, for example, \cite{thierry1980geometrical,bonora1983some}) is also simply a geometric fact in the algebroid context arising from the observation that the curvature of a Lie algebroid connection is zero when contracted with vertical vector fields. Working in the consistent splitting of the trivialized algebroid, this version of the Russian formula can be stated in a more familiar form as:
\begin{equation} \label{Algebroid Russian Formula 2}
	\Omega_\tau = \hatd_{\tau} \omega_\tau + \frac{1}{2}[\omega, \omega]_L = (\td + \ts)(b^A - \varpi^A)\otimes \un t_A + \frac{1}{2}[b-\varpi, b-\varpi]_{L} = \td b + \frac{1}{2}[b, b]_L = F\,,
\end{equation}
where $F \equiv \td b + \frac{1}{2}[b, b]_L$ is the gauge field strength of the gauge field $b$. In words, the curvature $\Omega_\tau$ is automatically ``ghost free'' without the need to apply any additional requirements.

In the BRST context, the Russian formula leads to the descent equations which subsequently characterize anomalies from a topological point of view \cite{alvarez1984topological,zumino1985chiral,zumino1984chiral,wess1971consequences}. This form of the anomaly is referred to as the \emph{consistent anomaly} as it satisfies the Wess-Zumino consistency condition. However, the consistent form of the anomaly is not gauge covariant, and one can separately introduce the corresponding covariantized version, called the \emph{covariant anomaly} \cite{bardeen1984consistent}. In this final section we will demonstrate how this story carries over into the algebroid language. Moreover, we will give an illustration of how the algebroid may afford us with a more complete picture by demonstrating that it is capable of geometrizing the consistent form of the anomaly as well as the covariant form. The conventional analysis of the BRST complex can only cover the former. Here we will be computing anomalies from a purely cohomological perspective which is independent of any particular physical theory. In other words, we simply mean that the consistent and covariant anomaly polynomials we derive have the correct topological and algebraic properties to be the anomalous divergences of the consistent and covariant currents that appear in the familiar physical considerations. 

\subsection{Characteristic Classes and Lie Algebroid Cohomology}

The cohomological formulation of anomalies begins by considering characteristic classes and their associated Chern-Simons forms. In this section we will work in the context of an arbitrary Atiyah Lie algebroid $A$, with connection reform $\omega$. Recall that the curvature of the connection reform is given by $\Omega = \hatd\omega + \frac{1}{2}[\omega,\omega]_L$. 

We begin by computing
\begin{equation}\label{Bianchi Identity}
\hatd\Omega = -[\omega,\Omega]_L\,,
\end{equation}
which can be recognized as the Bianchi identity, given $\hatd^2=0$. The pair of equations
\begin{equation}\label{Algebra of omega and Omega}
\hatd \omega = \Omega - \frac{1}{2}[\omega,\omega]_L, \qquad \hatd \Omega = -[\omega, \Omega]_L
\end{equation}
imply that the ring of polynomials generated by $\omega$ and $\Omega$ form a closed subalgebra of $\Omega(A)$. This is the basis of the Chern-Weil homomorphism, which states that one can formulate cohomology classes in $\Omega(A)$ using such polynomials \cite{WeilLetter,zbMATH03070474,chern1946characteristic,
chern1966geometry}.

To be precise, let $Q: L^{\otimes l} \rightarrow \mathbb{R}$ correspond to a symmetric, order-$l$ polynomial function on $L$ which is invariant under Lie algebroid morphisms. Such an object can be represented by a symmetric $l$-linear map in the tensor algebra of $L$. In other words, given the dual basis $\{t^A\}$ for $L^*$, with $A = 1, \ldots, \text{dim}(G)$, we can write $Q = Q_{A_1 \ldots A_l} \bigotimes_{j = 1}^l t^{A_j}$. In terms of such a symmetric, invariant polynomial we can define the \emph{characteristic class}
\begin{equation}\label{CC of symmetric invariant poly}
\lambda_Q(\Omega) = Q(\underbrace{\Omega, \ldots, \Omega}_l) = Q_{A_1 \ldots A_l} \wedge_{j = 1}^l \Omega^{A_j} \in \Omega^{2l}(A)\,.
\end{equation}
The Chern-Weil theorem\footnote{Strictly speaking, the Chern-Weil theorem is proven in the context of principal bundle cohomology. However, the basis of the proof hinges on the fact that the principal connection and curvature satisfy the same algebraic relations as the algebroid connection and curvature given in \eqref{Algebra of omega and Omega}. Hence, the proof carries over to this case as well. See \cite{fernandes2002lie} for a more rigorous discussion.} establishes that each $\lambda_Q(\Omega)$ defines an element of the cohomology class of degree $2l$ in the exterior algebra $\Omega(A)$. Specifically, it consists of the following two statements \cite{nakahara2018geometry}:
\begin{enumerate}
	\item Characteristic classes are closed $2l$-forms in $\Omega(A)$:
\begin{equation}
	\hatd \lambda_Q(\Omega) = l! Q(\hatd \Omega, \underbrace{\Omega, \ldots, \Omega}_{l-1}) = l! Q(\hatd \Omega + [\omega, \Omega]_{L}, \underbrace{\Omega, \ldots, \Omega}_{l-1}) = 0\,,
\end{equation}
which follows from the symmetry of $Q$ and the Bianchi identity. 
	\item Given two different connections $\omega_1$ and $\omega_2$, with respective curvatures $\Omega_1$ and $\Omega_2$, we have that $\lambda_Q(\Omega_2) - \lambda_Q(\Omega_1) \in \Omega^{2l}(A)$ is $\hatd$-exact. The relevant $(2l-1)$-form potential is defined by introducing a one parameter family of connections $\omega_t = \omega_1 + t(\omega_2 - \omega_1)$ which interpolates between $\omega_1$ and $\omega_2$ as $t$ goes from $0$ to $1$. Then,
\begin{equation} \label{Transgression Formula}
	\lambda_Q(\Omega_2) - \lambda_Q(\Omega_1) = \hatd \left[ Q_{A_1 \cdots A_l} \int_{0}^{1} \td t \; (\omega_2 - \omega_1)^{A_1} \wedge_{j = 2}^{l} \left(\hatd \omega_t + \frac{1}{2}[\omega_t , \omega_t]_{L}  \right)^{A_j} \right]\,.
\end{equation}
\end{enumerate}

An immediate corollary of the Chern-Weil theorem is that the characteristic class $\lambda_Q(\Omega)$ will be globally exact if there exists a one parameter family of connections for which $\omega_2 = \omega$ and $\omega_1$ is any connection that has zero curvature.\footnote{Note that a connection having zero curvature does not imply $\omega = 0$, which would be inconsistent with $\omega \circ j = -Id_L$. Rather, in the consistent splitting one can realize a connection with zero curvature by ensuring that the gauge field vanishes, i.e., $b = 0$. This implies $\omega_{\tau} = -\varpi$, which is consistent with the aforementioned identity. In physical contexts, this corresponds to the case that the connection is ``pure gauge".} This inspires the topological interpretation of the characteristic class which will be cohomologically trivial if and only if any connection $\omega$ can be homotopically connected to the trivial connection. Nonetheless, it will always be true locally that any characteristic class can be written as $\hatd$ acting on a $(2l-1)$-form defined using \eqref{Transgression Formula}. That is,
\begin{equation}\label{Chern Simons in algebroid}
	\lambda_Q(\Omega) = \hatd \mathscr{C}_Q(\omega)\,,
\end{equation} 
where
\begin{equation}
\label{transgression}
	\mathscr{C}_Q(\omega):= Q_{A_1 \cdots A_l} \int_{0}^{1} \td t\, \omega^{A_1} \wedge_{j = 2}^{l} \left(t\hatd \omega + \frac{1}{2}t^2[\omega , \omega]_{L}  \right)^{A_j}
\end{equation} 
is the Chern-Simons form associated with the symmetric invariant polynomial $Q$. Note that \eqref{Chern Simons in algebroid} indicates that there does not exist $\gamma\in\Omega^{2l-2}(A)$ such that $\scr C_Q=\hat\td\gamma$, and $\scr C_Q$ can only be determined up to a $\hat\td$ closed term. As we will see, a characteristic class $\lambda_Q(\Omega)$ and its associated Chern-Simons form $\mathscr{C}_Q(\omega)$ play central roles in the cohomological analysis of anomalies.

\subsection{Descent Equations and the Consistent Anomaly}
\label{sec:consistent}
Now, let us move into the trivialized algebroid and work in the consistent splitting. As we have shown, in the consistent splitting $\omega_\tau = b - \varpi$, and $\hatd_{\tau} \to \td + \ts$. It is therefore natural to organize the Chern-Simons form order by order in the bi-complex $\Omega(TM,L)$ as
\begin{equation} \label{Algebroid CS Expansion}
	\mathscr{C}_Q(b - \varpi) = \sum_{r + s = 2l-1} \alpha^{(r,s)}(b,\varpi)\,,
\end{equation}
where $\alpha^{(r,s)}(b,\varpi) \in \Omega^{(r,s)}(TM,L)$, and $\alpha^{(2l-2,1)}(b,\varpi) = \mathscr{C}_Q(b)$. 

Combining \eqref{Algebroid Russian Formula 2} and \eqref{Chern Simons in algebroid} yields
\begin{equation} \label{Algebroid Descent 1}
	\hatd_\tau \mathscr{C}_Q(b - \varpi) = \lambda_Q(\Omega) = \lambda_Q(F) = \td \mathscr{C}_Q(b)\,.
\end{equation} 
From this point it is straightforward to derive the descent equations simply by plugging (\ref{Algebroid CS Expansion}) into (\ref{Algebroid Descent 1}), and enforcing the equality order by order in the bi-complex $\Omega^{(r,s)}(TM,L)$. The descent equations can be expressed as
\begin{equation}\label{Descent Equations}
	\td \alpha^{(r,s)}(b,\varpi) + \ts\alpha^{(r+1,s-1)}(b,\varpi) = 0\,,\qquad r + s = 2l-1\,,\quad r \neq 2l-1\,,
\end{equation}
In particular, the term with $r = 2l-3$ yields the Wess-Zumino consistency condition:
\begin{equation} \label{WZalgebroid}
	\td\alpha^{(2l-3,2)}(b,\varpi) + \ts\alpha^{(2l-2,1)}(b,\varpi) = 0\,.
\end{equation}
On the other hand, from the fact that $\mathscr{C}_Q(b - \varpi)$ is not $\hat\td_\tau$ exact we also have
\be
\label{WZ not exact}
\alpha^{(2l-2,1)}(b,\varpi)\neq\td \gamma^{(2l-3,1)}(b,\varpi)+\ts\gamma^{(2l-2,0)}(b,\varpi)\,.
\ee
The term $\alpha^{(2l-2,1)}(b,\varpi)$ satisfying \eqref{WZalgebroid} and \eqref{WZ not exact} is a candidate to be the density of the consistent anomaly (see \cite{zumino1985chiral,zumino1984chiral,Gockeler:1987an} for a description from a physical and algebraic perspective). Thus, we have now demonstrated that the consistent anomaly arises naturally in the algebroid context:
\begin{equation}
 \label{Aconsistent}
	\mathfrak{a}_{\rm con} = \int_M \alpha^{(2l-2,1)}(b,\varpi)\,.
\end{equation}

\subsection{The Horizontal-Vertical Splitting and the Covariant Anomaly}
\label{sec:covariant}

Strictly speaking, the results discussed in the previous subsection 
are merely a reformulation of those obtained in the BRST analysis \cite{Bilal:2008qx}, although now they come from a transparent formal and geometric foundation which makes their origin and meaning clear. However, beyond simply improving our interpretation of the BRST analysis, we would now like to demonstrate that the algebroid approach has the potential to produce new results in the study of anomalies. 

As we have stressed, the trivialized algebroid has two relevant splittings. By analyzing the cohomology of the consistent splitting above we found the consistent anomaly. This inspires the question of whether the covariant splitting also has an interpretation related to an anomaly. Following the previous subsection, we can instead organize the Chern-Simons form on $A_{\tau}$ order by order in the bi-complex $\Omega^{(r,s)}(H_{\tau},V_{\tau})$. The most transparent way of doing this is by expanding the Chern-Simons form as a polynomial in the connection $\omega \in \Omega^1(V;L)$ and its curvature $\Omega \in \Omega^2(H;L)$. Here again we see the Russian formula playing a crucial role in dictating that the curvature can generate a sub-algebra of $\Omega(H_{\tau})$. The expansion of the Chern-Simons form can now be written as
\begin{equation}\label{covariantCSexpn}
	\mathscr{C}_Q(\omega) = \sum_{r + s = 2l-1} \beta^{(r,s)}(\omega,\Omega)\,,
\end{equation} 
where $\beta^{(r,s)}(\omega,\Omega) \in \Omega^{(r,s)}(H,V)$ contains $r/2$ factors of the curvature and $s$ factors of the connection.

We will now show that the covariant splitting directly produces the covariant anomaly. As was established in \cite{bardeen1984consistent,stone2012gravitational,Hughes:2012vg} the covariant anomaly is obtained from the free variation of the Chern-Simons form with respect to the connection. Computing this variation in the algebroid context, one arrives at the following formula (see Appendix~\ref{app:covanomaly} for details):
\begin{equation} 
\label{Covariant Anomaly}
	\delta \mathscr{C}_Q(\omega) = l \beta^{(2l-2,1)}(\delta\omega, \Omega) + \hatd\Theta(\omega,\delta\omega)\,,
\end{equation}
where
\begin{equation} 
\label{beta2l-2}
\beta^{(2l-2,1)}(\delta\omega, \Omega)=\frac{1}{l}Q(\underbrace{\Omega,\ldots,\Omega}_{l-1},\delta\omega)\,.
\end{equation}
Hence, the covariant anomaly can be read off from the first term in \eqref{Covariant Anomaly}. We therefore recognize that the covariant anomaly is intimately related to the term of order one in the vertical part of the Lie algebroid exterior algebra appearing in the expansion of the Chern-Simons form. This establishes a pleasant symmetry between the covariant anomaly and the consistent anomaly, since the consistent anomaly was proportional to the ``ghost number" one term in the expansion of the Chern-Simons form when viewed in the consistent splitting. We should note that from this point of view, the consistent and covariant anomalies do not coincide precisely because $V^*$ is not canonical, depending on the connection.

The covariant anomaly does not come with a series of descent equations that leads to a consistency condition. Instead, its defining property is that it is covariant with respect to the gauge transformation. In fact, we can now readily interpret the geometric difference between the consistent and covariant anomalies in the algebroid formulation. The former, being written in the consistent splitting of the algebroid, respects the nilpotency of the coboundary operator $\hatd$ in both factors of its associated bi-complex but spoils the gauge covariance. Conversely, the latter, although it does not admit two nilpotent differential operators, respects the covariant splitting defined by the connection $\omega$ and thus is endowed with  gauge covariance. Such a conclusion was not possible from the perspective of the BRST complex, precisely because it lacked a geometry for its connection to define a covariant splitting.

\subsection{Examples} \label{sec:examples}
We close this section by exploring a pair of illuminating examples, namely the chiral anomaly and the (type A) Lorentz-Weyl anomaly in $2d$. In both cases the covariant and consistent forms of the anomaly are deduced by analyzing an appropriate characteristic class and its associated Chern-Simons form. The analysis done here can easily be generalized to arbitrary even dimension.

\subsubsection{Chiral Anomaly in $2d$}\label{sec:chiral}
The analysis of the chiral anomaly arises in the context of an Atiyah Lie algebroid $A$ derived from a principal bundle $P(M,G)$, where $G$ is a semisimple Lie group. The characteristic class that is relevant to the chiral anomaly in $2d$ is the second Chern class\footnote{For simplicity, we have taken a basis such that the second Killing form is given by $\delta_{AB}$.}
\begin{equation}
\text{ch}_2(\Omega) = \delta_{AB} \; \Omega^A \wedge \Omega^B\,.
\end{equation}
The Chern-Simons form associated with $\text{ch}_2(\Omega)$ can be deduced by employing the transgression formula \eqref{Transgression Formula}:
\begin{equation}\label{CS form for second Chern class}
\mathscr{C}_2(\omega) = \delta_{AB}\left(\omega^A \wedge \hatd\omega^B + \frac{1}{3}\omega^A \wedge [\omega,\omega]_L^B\right)\,.
\end{equation}

Using \eqref{CS form for second Chern class}, we can easily determine the algebraic form of candidates for the covariant and consistent forms of the anomaly. To begin, still working in the algebroid $A$ we can decompose \eqref{CS form for second Chern class} order by order in the bi-complex $\Omega(H,V)$ by re-expressing it as a polynomial in the curvature and connection; that is, where there is a $\hatd\omega$ we will replace it by $\Omega - \frac{1}{2}[\omega,\omega]_L$. The resulting expression is
\begin{equation}
\mathscr{C}_2(\omega,\Omega) = \delta_{AB} \left(\omega^A \wedge \Omega^B - \frac{1}{6}\omega^A \wedge [\omega,\omega]_L^B\right)\,.
\end{equation}
In other words, the various terms in \eqref{covariantCSexpn} are given by
\begin{equation}
\beta^{(2,1)}(\omega,\Omega) = \delta_{AB} \; \omega^A \wedge \Omega^B\,, \qquad \beta^{(0,3)}(\omega,\Omega) = -\frac{1}{6}\delta_{AB} \; \omega^A \wedge [\omega,\omega]_L^B\,,
\end{equation}
from which we can read off  by applying \eqref{Covariant Anomaly} that the covariant anomaly polynomial is given in terms of the curvature $2\delta_{AB}\Omega^B$, as expected. 

To obtain the consistent anomaly polynomial, we pass to the trivialized Lie algebroid. That is, we specify a map $\tau: A \rightarrow A_{\tau}$ along with its inverse map $\overline{\tau}: A_{\tau} \rightarrow A$. Recall from Subsection \ref{sec:morphisms} that such a morphism implies the following relationships between the connections, curvatures, and coboundary operators of the two algebroids:
\begin{equation}
\overline{\tau}^*\omega = \omega_{\tau} = b - \varpi\,,\qquad 
\overline{\tau}^*\Omega = \Omega_{\tau} = F\,,\qquad 
\overline{\tau}^* \circ \hatd = \hatd_{\tau} \circ \overline{\tau}^* \,.
\end{equation}
Trivializing the Chern-Simons form, it follows from \eqref{hatdtauomega} that
\begin{eqnarray}\label{Trivialized CS Form for Chern Class}
\overline{\tau}^*\mathscr{C}_2(\omega)=\mathscr{C}_2(\omega_\tau)
=\mathscr{C}_2(b) + \delta_{AB} \left(-\varpi^A \wedge \td b^B - \frac{1}{2}b^A \wedge [\varpi,\varpi]_L^B + \frac{1}{6}\varpi^A \wedge [\varpi,\varpi]_L^B\right)\,.
\end{eqnarray}
Then, the expansion \eqref{Algebroid CS Expansion} gives
\begin{equation}
\begin{split}
\alpha^{(3,0)}(b,\varpi) &= \mathscr{C}_2(b)\,,\qquad 
\alpha^{(2,1)}(b,\varpi) = -\delta_{AB}  \varpi^A \wedge \td b^B\,,\\
\alpha^{(1,2)}(b,\varpi) &= -\frac{1}{2}\delta_{AB}  b^A \wedge[\varpi,\varpi]_L^B\,,\qquad
\alpha^{(0,3)}(b,\varpi) = \frac{1}{6}\delta_{AB}  \varpi^A \wedge [\varpi,\varpi]_L^B\,.
\end{split}
\end{equation}
The consistent anomaly polynomial can therefore be read off from the ghost number one contribution to \eqref{Trivialized CS Form for Chern Class}, which is $-\delta_{AB} \varpi^A \wedge \td b^B$. Recall that $-\varpi^A$ corresponds to the ghost field, the consistent anomaly can be recognized $\delta_{AB} \td b^B$, which is again in agreement with the known result. 

As promised, the covariant anomaly, which is written in terms of $\Omega$, is indeed covariant, while the consistent anomaly, which is written in terms of $\td b$, is not. Moreover, it is straightforward to show that the series of terms $\alpha^{(r,s)}(b,\varpi)$ satisfy the descent equations as introduced in \eqref{Descent Equations}.

\subsubsection{Lorentz-Weyl Anomaly in $2d$}\label{sec:LW}
To analyze the Lorentz-Weyl (LW) anomaly, let us begin by introducing the geometric framework and characteristic classes for a Lorentz-Weyl structure in arbitrary even dimension $d = 2l$. Consider an Atiyah Lie algebroid $A$ derived from a principal $G$-structure with $G = SO(1,d-1)\times\RR_+\subset GL(d,\RR)$. Here $SO(1,d-1)$ is the local Lorentz group, while $\RR_+$ corresponds to local Weyl rescaling. The corresponding Lie algebra can be expressed as $\mathfrak g=\mathfrak{so}(1,d-1)\oplus\mathfrak{r}_+$. The adjoint bundle of the group $G$ is given by $L = P \times_{G} \mg = L_{L} \oplus L_{W}$, where $L_L = P \times_{SO(1,d-1)} \mathfrak{so}(1,d-1)$ and $L_{W} = P \times_{\mathbb{R}_+} \mathfrak{r}_+d$ correspond to the Lorentz and Weyl factors, respectively. 
The connection reform on $A$ will therefore split as $\omega =\omega_L + \omega_W$ where $\omega_L$ and $\omega_W$ are the connection reform on the Lorentz and Weyl sub-algebroids, respectively. The curvature of the connection reform $\omega$ will have two pieces
\begin{equation}
\Omega = \hatd\omega + \frac{1}{2}[\omega,\omega]_L = \Omega_{L} + \Omega_{W}\,,
\end{equation}
where $\Omega_{L}\in \Omega^2(H;L_L)$ is related to the Riemann tensor and $\Omega_{W} \in \Omega^2(H;L_{W})$ is the gauge field strength of the Weyl connection. We can see that the curvature $\Omega$ remains horizontal.

There are two natural invariant structures associated with $L$. The Weyl factor $L_{W}$ is an Abelian subalgebra of $L$. Thus, the map $\tr_W: L \rightarrow L_{W}$ which projects an element $\underline{\mu} \in \Gamma(L)$ down to $L_{W}$ will be invariant under the adjoint action of $L$ on itself. In a linear representation of $L$ given by $v_E: L \rightarrow \text{End}(L)$, the generators of $L_L$ are represented by traceless antisymmetric matrices. Hence, as the notation indicates, the map $\tr_W$ can also be understood by selecting a representation and computing the ordinary trace. In other words, for any representation $E$ and given $\tr: \text{End}(E) \rightarrow C^\infty(M)$ we have
\begin{equation} \label{Trace Map}
\tr_W(\underline{\mu}) = \tr \circ v_E(\underline{\mu})\,.
\end{equation} 

Similarly, there is an invariant structure on $L_L$ which will correspond to the Pfaffian. In particular we define
\begin{equation} \label{Pfaffian Map}
\epsilon: L^{\otimes l} \rightarrow C^{\infty}(M)\,.
\end{equation}
One of the defining properties of the map $\epsilon$ is that $\epsilon(\underline{\mu}_1, \ldots , \underline{\mu}_{l}) = 0$ if $\underline{\mu_i} \in \Gamma(L_{W})$ for any $i$. In other words, $\epsilon$ only sees the orthogonal factor of $G$, and is an invariant polynomial on this factor. As was the case with the trace, $\epsilon$ can be computed by passing to a linear representation. To be precise, we should take a $2l$-dimensional representation space $E$ equipped with an inner product $g_E: E \times E \rightarrow C^{\infty}(M)$ of appropriate signature. Then, we can define the map $w_E: L \rightarrow \wedge^2 E^*$ such that given $\underline{\psi}_1, \underline{\psi}_2 \in \Gamma(E)$ we have
\begin{equation}
w_E(\underline{\mu})(\underline{\psi}_1, \underline{\psi}_2) = g_E\left(\underline{\psi}_1, v_E(\underline{\mu})(\underline{\psi}_2)\right)\,.
\end{equation}
Notice that $w_E \circ \tr_W = 0$, since a Weyl rescaling cannot be represented by an antisymmetric matrix. Given an oriented orthonormal basis $\{\underline{e}_a\}$ for $E$ along with its dual basis $\{e^a\}$, with $a=1,\ldots, 2l$, we can define an $SO(1,d-1)$ invariant volume form on $E$\footnote{Note that we are {\it not} specifying a solder form, and so we have no way to pull this volume form back to the base. Similarly the inner product on $E$ is not directly related to a metric on the base. These facts might be thought of as being responsible for the topological nature of the characteristic classes discussed below.}
\begin{equation}
\text{Vol}_E \equiv \epsilon_{a_1 \cdots a_d} e^{a_1} \wedge \cdots \wedge e^{a_d}\,.
\end{equation}
Thus, in this representation we can express:
\begin{equation}\label{Evaluation of Pffafian Map}
\epsilon(\underline{\mu}_1, \ldots, \underline{\mu}_l) = \epsilon_{a_1 b_1 \cdots a_l b_l} w_E(\underline{\mu}_1)^{a_1 b_1} \cdots w_E(\underline{\mu}_l)^{a_l b_l} = \epsilon^{a_1}{}_{b_1} \cdots^{a_l}{}_{b_l} v_E(\underline{\mu}_1)^{b_1}{}_{a_1} \cdots v_E(\underline{\mu}_l)^{b_l}{}_{a_l}\,.
\end{equation}
This construction satisfies the above-mentioned properties since $w_E \circ \tr_W(\underline{\mu}) = 0$ and 
\begin{equation}
\epsilon(\underline{\mu}, \ldots, \underline{\mu}) = \text{Pf}(\underline{\mu})\,.
\end{equation}
Note that this construction requires $d$ to be even, as the $\epsilon^{a_1}{}_{b_1} \cdots^{a_l}{}_{b_l}$ has an equal number of up and down indices (signifying its Weyl invariance).

We are now prepared to introduce the relevant characteristic class for the LW anomaly. If we intend to derive the anomaly for a $d = 2l$ dimensional theory, we must construct a characteristic class of form degree $d+2 = 2(l+1)$. Hence, we must construct a symmetric and invariant linear map $Q^{LW,l+1}: L^{\otimes (l+1)} \rightarrow \mathbb{R}$. As we have discussed, we have at our disposal two invariant objects corresponding to the trace \eqref{Trace Map} and the Pfaffian \eqref{Pfaffian Map}. We therefore obtain an $(l+1)$-order symmetric invariant polynomial by taking the symmetrized product of these two maps:
\begin{equation} \label{symminv}
Q^{LW,l+1}(\underline{\mu}_1, \ldots \underline{\mu}_{l+1}) = \sum_{\pi} \epsilon(\underline{\mu}_{\pi(1)}, \ldots, \underline{\mu}_{\pi(l)})\,\tr_{W}(\underline{\mu}_{\pi(l+1)})\,,
\end{equation}
where $\pi$ denotes the permutations of $(1,\ldots,l+1)$. The characteristic class associated with $Q^{LW,l+1}$ is therefore given by $\lambda_{Q^{LW,l+1}}(\Omega)$ as dictated in \eqref{CC of symmetric invariant poly}.  While $\lambda_{Q^{LW,l+1}}$ is the appropriate characteristic class in the LW context, in other situations (such as a simple or semi-simple group) one finds an Euler class.\footnote{Indeed in the literature \cite{Boulanger:2007st,Boulanger:2018rxo,Francois:2015oca,Francois:2015pg} there is an analysis of Cartan geometry, in which the symmetry is enhanced to $SO(2,d)$, and the type A conformal anomaly comes from the Euler class. Descending to the subgroup $SO(1,d-1)\times\RR_+$ considered here, one obtains \eqref{symminv}.}

Let us now specialize to the case $d = 2$ and show that $\lambda_{Q^{LW,2}}$ gives rise to the LW anomaly. The characteristic class of interest takes the following form:
\begin{equation}\label{Gen 2 Euler}
\lambda_{Q^{LW,2}}(\Omega) = \frac{1}{2}\left(\epsilon(\Omega) \wedge \tr_W(\Omega) + \tr_W(\Omega) \wedge \epsilon(\Omega)\right)\,.
\end{equation}
In the $2d$ case, since the structure group $G = SO(1,1) \times \mathbb{R}_{+}$ is Abelian, we can write $\Omega = \hatd\omega$. Hence, the Chern-Simons form can be obtained as
\begin{equation}\label{CS form for second generalized euler class}
\mathscr{C}_{LW,2}(\omega,\Omega) = \frac{1}{2}\left(\epsilon(\omega) \wedge \tr_W(\Omega) + \tr_W(\omega) \wedge \epsilon(\Omega)\right)\,.
\end{equation}
To read off the covariant form of the anomaly polynomial let us pass to a representation on $E$. Then using \eqref{Trace Map} and \eqref{Evaluation of Pffafian Map} we can write the covariant anomaly as (ignoring the constant factor)
\begin{equation}
\label{LW cov anom}
\Omega_{W}\epsilon^{a}{}_b + \text{Pf}(\Omega_{L})\delta^{a}{}_b\,.
\end{equation}
Noticing that $\epsilon(\omega)$ and $\tr_W(\omega)$ picks out the Lorentz and Weyl part of the connection, respectively, the first term in the above result should be interpreted as the Lorentz anomaly, which vanishes when the Weyl connection is turned off; the second term is the Weyl anomaly in $2d$, which is proportional to the Ricci scalar of the spacetime. Therefore, the LW anomaly is the mixed anomaly between the Lorentz and Weyl symmetry. In fact, it is easy to see that by adding a total derivative term, one can remove the Lorentz anomaly or Weyl anomaly but cannot remove both simultaneously.

To obtain the consistent form, we must employ a Lie algebroid trivialization. Under the trivialization we find that
\begin{equation}
\overline{\tau}^*\omega = b - \varpi_{L} + a - \varpi_{W}\,,\qquad 
\overline{\tau}^*\Omega = R + f\,,\qquad 
\overline{\tau}^* \circ \hatd = (\td + \ts_{L} + \ts_{W}) \circ \overline{\tau}^*\,,
\end{equation}
where $b$ and $a$ are the spin connection and Weyl connection on $M$, and $R$ and $f$ are their curvature 2-forms, respectively. The pairs $(\varpi_{L},\ts_{L})$ and $(\varpi_{W},\ts_{W})$ are the Maurer-Cartan forms and BRST operators for the $SO(1,1)$ and $\mathbb{R}_+$ factors of $L$. Let $B = b + a$ and $\varpi = \varpi_{L} + \varpi_{W}$ denote the combined gauge field and Maurer-Cartan forms. We subsequently identify the consistent LW anomaly from $Q^{LW,2}(\varpi,\td B)$. Since in the index notation of the representation we have 
\be
(\td B)^a{}_b = R\epsilon^a{}_b + f\delta^a{}_b\,,
\ee
the consistent form of the LW anomaly is merely the pullback of the covariant form by the trivialization $\overline{\tau}$, which reads
\begin{equation}
f\epsilon^{a}{}_b + \text{Pf}(R)\delta^{a}{}_b\,,
\end{equation}
which has the same form as \eqref{LW cov anom}. This follows in this particular case from the fact that $G$ is an Abelian group when $d=2$.

A simplified account of the LW anomaly in two dimensions appeared recently in Appendix A of \cite{Campoleoni:2022wmf}.
Note that here we have focussed on the type A Weyl anomaly, and the type B Weyl anomaly remains an open question in general dimension. A more elaborate discussion is required since obstruction tensors are expected to make an appearance 
\cite{fefferman2001q,Graham2005ambient,graham2009extended,jia2021obstruction,Jia:2023gmk}. 
We expect to return to this issue, as well as other $G$-structures, in a future publication.

\section{Conclusions} \label{sec:disc}
In the introduction we raised a series of questions about the BRST formalism. In the course of this paper we have provided answers to each of these questions by geometrically formalizing the BRST complex in terms of the Atiyah Lie algebroid. As we promised in the introduction, each answer follows immediately from the geometry of the Atiyah Lie algebroid.

\textbf{Q:} Why should the Grassmann-valued fields $c^A(x)$, which started their life in the BRST quantization procedure have an interpretation as the generators of local gauge transformations? And why is it reasonable to combine the de Rham complex and the ghost algebra into a single exterior bi-algebra?

\textbf{A:} In the algebroid context the Maurer-Cartan form $\varpi \in \Omega^1(L;L)$ plays the role of the gauge ghost, and is also a generator of local gauge transformations. Working in the consistent splitting the exterior algebra of the trivialized algebroid $A_{\tau}$ subsequently takes the form of a bi-complex $\Omega^{(p,q)}(TM,L;E)$, where $p$ is the form degree with respect to the de Rham cohomology of $M$, and $q$ is the ``ghost number". The coboundary operator $\hat{\td}_{\tau}$ takes explicitly the form $\td + \ts$  on this exterior algebra, where $\td$ is the de Rham differential and $\ts$ is the BRST operator. 

\textbf{Q:} Why is it reasonable to consider $\econn = A + c$ as a ``connection'', and moreover what horizontal distribution does it define?

\textbf{A:} Still in the context of the trivialized Lie algebroid, one can introduce a connection reform, $\omega_\tau: A_{\tau} \rightarrow L$, defining the horizontal distribution $H_{\tau} = \text{ker}(\omega_\tau)$ for which $A_{\tau} = H_{\tau} \oplus V_{\tau}$. In the consistent splitting $\omega_\tau = b - \varpi$, where $b: TM \rightarrow L$ is a local gauge field, and $\varpi: L \rightarrow L$ is the Maurer-Cartan form on $L$. Hence, $\omega$ reproduces the ``connection" $\econn$ defined in the BRST complex, where again we see the role of the gauge ghost being played by the Maurer-Cartan form. 

\textbf{Q:} Why should the ``curvature" $\ecurv$ be taken to have ghost number zero? And why does enforcing this requirement turn the BRST operator $s$ into the Chevalley-Eilenberg operator for the Lie algebra of the structure group?

\textbf{A:} $\ecurv$ in the context of the trivialized Lie algebroid is represented by $\Omega_\tau = \hat{\td}_\tau\omega_\tau + \frac{1}{2}[\omega_\tau, \omega_\tau]_{L}$, namely the curvature associated with $\omega_\tau$, which is fully horizontal as a built-in geometric property of the algebroid. In the consistent splitting, this reproduces the Russian formula and the BRST transformation as we presented in \eqref{Algebroid Russian Formula 2}. 

The culmination of all of these facts gives rise to the descent equations \eqref{Descent Equations} and the Wess-Zumino consistency condition \eqref{WZalgebroid}. Given a characteristic class $\lambda_{Q}(\Omega)$ with associated Chern-Simons form $\mathscr{C}_{Q}(\omega)$ we have
\begin{equation}
    \hat{\td}_{\tau} \mathscr{C}_{Q}(\omega) = (\td + \ts) \mathscr{C}_{Q}(b - \varpi) = \td \mathscr{C}_{Q}(b)\,.
\end{equation}
From the above equation, one can immediately compute the \emph{consistent} anomaly polynomial, which corresponds to the ghost number one contribution to $\mathscr{C}_Q(b - \varpi)$, and can be shown to be an element of the first cohomology of the BRST operator $\ts$ once integrated over a space of appropriate dimension. Furthermore, one can also obtain the \emph{covariant} form of the anomaly by viewing the Chern-Simons form in the covariant splitting and extracting the terms contributing with one exterior power in the vertical sub-bundle of the associated exterior algebra (multiplied by the order $l$ of $Q$). Although the formulae for finding the consistent and covariant anomalies have been known \cite{bardeen1984consistent}, our approach to these anomalies provides a meaningful explanation as to why the consistent anomaly is consistent and the covariant anomaly is covariant. From the algebroid perspective, they just correspond to different choices of splitting.

To understand the complete picture of the consistent and covariant anomalies as well as the anomaly inflow mechanism that relates them, we will have to further exploit the structure of the configuration space of Lie algebroid connections. In this paper we established a powerful approach for studying Lie algebroid isomorphisms in terms of commutative diagrams, which found a physical interpretation as a unified tool for implementing diffeomorphisms and gauge transformations. In a partner paper \cite{klinger2023abc} we make use of this construction to define a new geometric formalism for understanding the extended configuration space of arbitrary gauge theories. We refer to this construction as the configuration algebroid. We demonstrated that the configuration algebroid provides a suitable quantification of the local degrees of freedom in a gauge theory, leading to a fully integrable algebra of charges associated with the local symmetries of a theory. From the point of view of the configuration algebroid, the presence of anomalies is associated with the question of whether the charge algebra is centrally extended. In forthcoming work we will combine the insights of this paper with \cite{klinger2023abc} to describe anomalies as topological features of the configuration algebroid, and demonstrate how the anomaly inflow mechanism can be incorporated into the algebroid language.

\section*{Acknowledgements}

We thank Luca Ciambelli, Pin-Chun Pai, Manthos Karydas and Mike Stone for conversations. This work was supported by the U.S. Department of Energy under contract DE-SC0015655.

\appendix
\renewcommand{\theequation}{\thesection.\arabic{equation}}
\setcounter{equation}{0}

\section{Chain Maps}
\label{app:chainmap}

In this appendix we offer a direct computation verifying that a Lie algebroid isomorphism $\varphi: A_1 \rightarrow A_2$ satisfying the compatibility condition \eqref{Iso-Rep compatibility} induces a chain map on the exterior algebras of $A_1$ and $A_2$. It is sufficient to show that this condition holds for 0-forms and 1-forms, since $\hatd$ acts as a derivation with respect to the wedge product and the full exterior algebra is generated by the set of 1-forms along with the wedge product. First we look at the 0-form case. Let $\psi \in \Omega^0(A_2;E_2)$, and $\un\mX\in\Gamma(A_1)$. Then,
\begin{align}
&(\varphi^* \hatd_2 \psi)(\un\mX) = g_{\varphi}^{-1}\big(\hatd_2\psi \circ \varphi(\un\mX)\big) = g_{\varphi}^{-1}\big(\Aconn{E_2} \circ \varphi(\un\mX)(\psi)\big) \nn \\
={}& g_{\varphi}^{-1}\big(\Aconn{E_2} \circ \varphi(\un\mX)\big(g_{\varphi}g_{\varphi}^{-1}(\psi)\big)\big) = \Aconn{E_1}(\un\mX)\big(g_{\varphi}^{-1}(\psi)\big) = (\hatd_1 \varphi^*\psi)(\un\mX)\,,
\end{align}
where in the first equality we used \eqref{Lie algebroid pullback}, in the second equality we used the definition of the Lie algebroid differential via the Koszul formula \eqref{dhat on E}, and in the fourth equatlity we used \eqref{Iso-Rep compatibility}. 
\par
Now we move on to the 1-form case. Let $\eta\in\Omega^1(A_2;E_2)$, and take $\un\mX,\un\mY\in\Gamma(A_1)$. We can write
\begin{align}
&(\varphi^*\hat\td_2\eta)(\un\mX,\un\mY)=g_{\varphi}^{-1}[(\hat\td_2\eta)(\varphi(\un\mX),\varphi(\un\mY))]\nn\\
={}& g_{\varphi}^{-1}\Big[\Aconn{E_2}\circ \varphi(\un\mX) (\eta \circ \varphi(\un\mY)) - \Aconn{E_2} \circ \varphi(\un\mY) (\eta \circ \varphi(\un\mX)) - \eta([\varphi(\un\mX),\varphi(\un\mY)]_{A_2})  \Big] \nn \\
={}& \Aconn{E_1}(\un\mX)\big(\varphi^*\eta(\un\mY)\big) - \Aconn{E_1}(\un\mY)\big(\varphi^*\eta(\un\mX)\big) - \varphi^*\eta\big([\un\mX,\un\mY]_{A_1}\big) \nn \\
={}&(\hatd_1\varphi^*\un\eta)(\un\mX,\un\mY)\,,
\end{align}
where again in the first equaltiy we used \eqref{Lie algebroid pullback}, in the second equality we used \eqref{dhat on E}, and in third equality we applied \eqref{Iso-Rep compatibility} and \eqref{Lie algebroid pullback}.

\section{Free Variation and the Covariant Anomaly}
\label{app:covanomaly}
In Subsection \ref{sec:covariant}, we introduced that the covariant anomaly can be derived by taking the free variation of the Chern-Simons form $\scr C_Q(\omega)$ in the covariant splitting, as shown in equation \eqref{Covariant Anomaly}. We will now provide an explicit demonstration of this derivation. Following the approach presented in \cite{bardeen1984consistent}, we introduce a nilpotent operator $K:\Omega^{p}(A;L)\to\Omega^{p-1}(A;L)$ that acts as follows:
\begin{align}
K\omega=0\,,\qquad K\Omega=\delta\omega\,,\qquad K\delta\omega=0\,.
\end{align}
Then, the variation operator on $\omega$ and $\Omega$ can be written as
\begin{align}
\delta=K\hat\td+\hat\td K\,.
\end{align}
When performing the variation of the Chern-Simons form:
\begin{align}
\label{KddKC}
\delta\mathscr{C}_Q=K\hat\td\mathscr{C}_Q+\hat\td K\mathscr{C}_Q\,,
\end{align}
the second term is a total derivative, and thus all we have to show is that the first term in \eqref{KddKC} gives rise to the first term in \eqref{Covariant Anomaly}, namely $\beta^{(2l-2,1)}(\delta\omega, \Omega)$. Using the transgression formula \eqref{transgression}, one finds
\begin{align}
K\hat\td\mathscr{C}_Q(\omega)={}&l Q_{A_1 \cdots A_l} \int_{0}^{1} \td t\,\delta\omega^{A_1} \wedge_{j = 2}^{l} \left(t\Omega + \frac{1}{2}(t^2-t)[\omega , \omega]_{L}  \right)^{A_j}\nn\\
\label{KdC1}
&+\frac{l-1}{2}Q_{A_1 \cdots A_l} \int_{0}^{1} \td t\,t^2\delta\omega^{A_1} [\omega , \omega]_{L}^{A_2}\wedge_{j = 3}^{l} \left(t\Omega + \frac{1}{2}(t^2-t)[\omega , \omega]_{L}  \right)^{A_j}\,.
\end{align}
To further evaluate this, it is not difficult to first perform the integral of the following form:
\begin{align}
\int_{0}^{1} \td t \left[ l\Big(t A+\frac{t^2-t}{2}B\Big)^{l-1}+\frac{l-1}{2}t^2 B\Big(t A+\frac{t^2-t}{2}B\Big)^{l-2} \right] =A^{l-1}\,.
\end{align}
Then, taking $A$ as $\Omega$ and $B$ as $[\omega,\omega]_L$, the integral in \eqref{KdC1} yields 
\begin{align}
K\hat\td\mathscr{C}_Q(\omega)=Q(\underbrace{\Omega,\ldots,\Omega}_{l-1},\delta\omega)\,.
\end{align}
Now we can compare this with $\beta^{(2l-2,1)}(\delta\omega, \Omega)$. From \eqref{transgression}, one can pick up the term with a single $\omega$ and find
\begin{align}
 \beta^{(2l-2,1)}(\omega, \Omega)=Q_{A_1 \cdots A_l} \int_{0}^{1} \td t\,\omega^{A_1} t^{l-1}\wedge_{j = 2}^{l} \Omega^{A_j}=\frac{1}{l}Q(\underbrace{\Omega,\ldots,\Omega}_{l-1},\omega)\,,
\end{align}
and hence
\begin{equation} 
\label{beta2l-2a}
\beta^{(2l-2,1)}(\delta\omega, \Omega)= \frac{1}{l}Q(\underbrace{\Omega,\ldots,\Omega}_{l-1},\delta\omega)\,.
\end{equation}
Therefore, we can see that \eqref{KddKC} can be written as
\begin{equation} 
\label{deltaC}
\delta \mathscr{C}_Q(\omega) = l \beta^{(2l-2,1)}(\delta\omega, \Omega) + \hat\td\Theta(\omega,\delta\omega)\,,
\end{equation}
where $\Theta\equiv K\mathscr{C}_Q$. The covariant anomaly can be read off from the first term, while the $\Theta$ in the second term serves as the Bardeen-Zumino polynomial which covariantizes the consistent anomaly when added to the anomalous current \cite{bardeen1984consistent}.

\providecommand{\href}[2]{#2}\begingroup\raggedright\endgroup

\end{document}